\documentclass{PoS}

\title{Benchmark values for the net proton number fluctuations}

\ShortTitle{Benchmark values for the net proton number fluctuations}

\author{\speaker{Boris Tom\'a\v{s}ik}\\
        Fakulta pr\'irodn\'ych vied, Univerzita Mateja Bela, Bansk\'a Bystrica, Slovakia\\and\\
        Fakulta jadern\'a a fyzik\'aln\v{e} in\v{z}en\'yrsk\'a, \v{C}esk\'e vysok\'e u\v{c}en\'i technick\'e v Praze, Prague, Czechia\\
        E-mail: \email{boris.tomasik@cern.ch}}

\author{Ivan Melo\\
        Fakulta elektrotechniky a informa\v{c}n\'ych technol\'ogi\'i, \v{Z}ilinsk\'a univerzita, \v{Z}ilina, Slovakia
}

\author{Luk\'a\v{s} Laff\'ers\\
Fakulta pr\'irodn\'ych vied, Univerzita Mateja Bela, Bansk\'a Bystrica, Slovakia
}

\author{Marcus Bleicher\\
Frankfurt Institute for Advanced Studies, 
Johann Wolfgang Goethe-Universit\"at,\\  Frankfurt am Main, Germany,
\\and\\
Institut f\"ur Theoretische Physik, 
Johann Wolfgang Goethe-Universit\"at,\\  Frankfurt am Main, Germany
}

\abstract{%
We formulate a baseline model for the net-proton number fluctuations
in ultrarelativistic heavy-ion collisions. Our model includes total baryon number 
conservation, fluctuations of the participating nucleon number, limited acceptance, and a possibility that isospin is 
remembered by the participating wounded nucleons. 
Most importantly, we formulated the model as consisting of two components: wounded nucleons and the produced 
nucleon-antinucleon pairs. Those two components have different rapidity distributions. Owing to this feature we obtain 
also predictions for the rapidity dependence of the cumulants of the net-proton number distribution and their ratios.
}

\FullConference{Corfu Summer Institute 2018 "School and Workshops on Elementary Particle Physics and Gravity"\\
		(CORFU2018)\\
		31 August - 28 September, 2018\\
		Corfu, Greece}

\begin{document}

\section{Introduction}

Statistical physics arguments lead to the conclusions about strongly interacting matter, 
that in the vicinity of the critical point of the phase diagram the susceptibilities 
of higher orders should strongly depend on temperature and the baryon chemical potential $\mu_B$ 
\cite{Hatta:2003wn,Athanasiou:2010kw,Gavai:2010zn,Stephanov:2011pb,Gupta:2011wh,Stephanov:2008qz}.
The susceptibilities with respect to $\mu_B$ represent themselves in the cumulants of the net-baryon number 
distribution \cite{Athanasiou:2010kw,Stephanov:2008qz}. 
In principle, they can be even calculated from the first principles by simulating the QCD partition function on the 
lattice \cite{Bazavov:2017tot,Borsanyi:2018grb}. 
These arguments and calculations are intrinsically connected with the grand-canonical ensemble of systems. 

Unfortunately, the baryon number fluctuations, as they are elegantly treated in the grand-canonical formalism, cannot be directly measured 
in an experiment. Three major reasons make this impossible: 
\begin{itemize}
\item 
In real collisions, baryon number is always conserved. Hence, if the whole phase-space would be covered by detectors that would 
register all baryons and antibaryons with perfect efficiency, there would be no fluctuations. On the other hand,
if the detectors cover only a part of the phase-space, then fluctuations may arise just from the binomial distribution of the 
appearance of a particle in the acceptance window \cite{Braun-Munzinger:2016yjz,Bzdak:2017ltv}. 
\item 
It is experimentally impossible to investigate a set of heavy-ion 
collisions with always exactly the same number of nucleons participating 
in the collisions. This brings in another source of baryon number fluctuations. It has to be taken into account also in theoretical 
simulations which are compared to experimental data. Another possibility is to filter out such fluctuations from the experimental data 
my means of unfolding \cite{Luo:2013bmi,Sugiura:2019toh}. 
\item 
Not all baryons (and antibaryons) can be detected. A typical detector does not register neutrons. Therefore, the net-proton  number  
is usually measured as a good proxy for the net baryon number. The argument in favour of this proxy is usually based on the high 
rate of the isospin-changing reactions that can change protons into neutrons and vice versa \cite{KA,KA2}.   Note, however, 
that such arguments may be of limited validity at the lowest collision energies within the RHIC Beam Energy Scan programme. 
\end{itemize}

The less-than-perfect acceptance of the detectors also acts as a source of fluctuations. This problem has been addressed in the 
literature and we will not analyse it here  \cite{Bzdak:2013pha,Luo:2014rea,Kitazawa:2016awu,Bzdak:2016qdc}. 
Instead, we assume that the detector efficiency is 100\%.

In this study we investigate the influence of the effects mentioned above on the net-baryon and the net-proton number 
fluctuations. Two new points are introduced in our treatment. 
\begin{itemize}
\item 
We simulate the participating baryons and the produced $B\bar B$ pairs with rapidity distributions that depend on the collision energy. 
This, combined with the rapidity window fixed for all collision energies (since it is really fixed by the actual detector), leads to the
collision energy dependence of the fluctuations. The rapidity distributions are constructed with an eye on the experimental data \cite{star2009,starbulk}. 
The distribution of the  wounded nucleons is different from the distribution of the produced $B\bar B$ pairs. 
\item 
Since for lower collision energies the isospin randomisation may become ineffective, we also explore the possibility 
that the  wounded nucleons remember their isospin. 
\end{itemize}

We study the fluctuations as functions of the size of the acceptance window in rapidity but also as a function 
of the position of the acceptance window. The latter is relevant as a baseline study for the recent proposal 
to look for the position of the critical point by measuring fluctuations at slightly forward rapidities, where the 
net baryon density may be higher than at midrapidity \cite{Brewer}. This is doable for us owing to the implemented rapidity 
distributions of the wounded nucleons.


\section{The model}

Our results will be based on Monte Carlo simulations that will assign the baryons and antibaryons their rapidity. 
First of all, we respect the baryon number conservation. Only protons, neutrons, and their antiparticles 
are included into our simulations. From the colliding gold nuclei only a part of the incoming nucleons 
participate in the collision. 

The number of wounded nucleons is determined with the help of GLISSANDO \cite{glis,glis2,glis1}. We first create a sample 
of minimum bias collisions for each of the investigated collision energies. 
Then we select centrality classes based
on the relative deposited strength (RDS) which should be proportional to the produced multiplicity
\begin{equation}
\label{e:rds}
M \propto \frac{1-\alpha}{2} N_w + \alpha N_{bin}\,  ,
\end{equation}
where the parameter $\alpha$ increases with the collision energy. It has been set by fitting the centrality dependence of the 
multiplicity in Au+Au collisions at $\sqrt{s_{NN}} = 19.6$~GeV and 200~GeV and interpolating for other energies 
with the logarithmic dependence
\begin{equation}
\alpha(\sqrt{s_{NN}}) = \alpha_0 + \alpha_1\ln\sqrt{s_{NN}}\,  .
\end{equation}
This method is  closest to what is done in real experiment. 

One of the features of our model, which can be turned on and off, is the isospin memory of wounded nucleons. 
This may be (partially) present in  collisions at the lowest energies. The isospin of a wounded nucleon is set in such a way 
that one never can obtain more protons (or neutrons) than there are in the incoming nucleus. This means that if 
$N_p$ wounded protons and $N_n$ wounded neutrons have been generated, then the probability that the next 
wounded nucleon will be a proton is $(Z-N_p)/(A-N_p-N_n)$, where $A$ and $Z$ refer to gold nucleus. The number of 
wounded protons is then distributed according to hypergeometric distribution. 

In addition to the wounded nucleons there are also produced pairs of nucleons and antinucleons. The number of pairs fluctuates 
according to Poisson distribution, with its mean proportional to the number of wounded nucleons
\begin{equation}
\mu_{B\bar B}= \frac{dN_{\bar p}}{dy} \, y_m\, \frac{N_w}{\langle N_w \rangle}\,  ,
\end{equation}
where $dN_{\bar p}/dy$ is the rapidity density of antiprotons measured at given energy and centrality \cite{star2009,starbulk}, 
$y_m$ is the width of 
the whole rapidity distribution of the produced $B\bar B$ pairs, $N_w$ is the number of wounded nucleons in the particular 
event, and $\langle N_w \rangle$ is the mean number of wounded nucleons at given collision energy and centrality, as determined 
by GLISSANDO. 
The isospin is assigned with equal probabilities for (anti)protons and (anti)neutrons. 
Antinucleons are, of course, only present among the produced $B\bar B$ pairs. 

Not the whole rapidity region is covered by the detectors, thus only a part of the (anti)nucleons with rapidities in the experimentally accepted
window is assumed to be registered. 

It is a crucial feature of this model, that the relative abundance of wounded and produced nucleons depends on the collision energy, centrality,
and the acceptance window in rapidity.

Finally, let us specify the rapidity distributions. Wounded nucleons partially remember their original rapidity, so we parametrize
their rapidity distribution with a double Gaussian function
\begin{equation}
\frac{dN_w}{dy} = \frac{N_w}{2\sqrt{2\pi \sigma_y^2}} \left \{ 
\exp\left ( - \frac{(y + y_m)^2}{2\sigma_y^2}\right )
+
\exp\left ( - \frac{(y - y_m)^2}{2\sigma_y^2}\right )
\right \}\,  .
\end{equation}
We set the width $\sigma_y = 0.8$, and $N_w$ is the total number of wounded nucleons in the given event. 
The shifts $y_m$ are determined from data. We use the fact that wounded protons make up the difference between 
observed protons and antiprotons. Thus the number of the observed net protons in a given rapidity window $(-y_b,y_b)$
can be determined
\begin{equation}
N_{p-\bar p} = \frac{Z}{A} \int_{-y_b}^{y_b} \frac{dN_w}{dy} \, dy\,  .
\end{equation}
We use data for $y_b = 0.25$ measured at different collision energies by the STAR collaboration \cite{star2009,starbulk}. 

For the produced nucleon-antinucleon pairs we assume that they are produced mainly around midrapidity. To soften sharp edges 
of their distribution we use the Woods-Saxon distribution
\begin{equation}
\frac{dN_{B\bar B}}{dy} = N_{B\bar B} \frac{C}{1 + \exp \left ( \frac{|y| - y_m}{a} \right )}\,  ,
\end{equation}
where $C$ is the proper normalisation constant $C = (2a \ln(e^{y_m/a} + 1))^{-1}$, and we choose $a = \sigma_y /10$.
The number of pairs $N_{B\bar B}$ is determined from data: note that this is the only source of antiprotons in our model. 
Thus we set the parameters so that we can reproduce the measured number of antiprotons $N_{\bar p}$ in the interval 
$|y| < y_b = 0.25$ \cite{star2009,starbulk}
\begin{equation}
N_{\bar p} = \frac{1}{2} \int_{-y_b}^{y_b} \frac{dN_{B\bar B}}{dy}\, dy\,  ,
\end{equation}
where the factor 1/2 in front of the integral stands for taking only antiprotons and no antineutrons. 

The pairs of nucleons and antinucleons are generated so that their mutual distances in rapidity are distributed exponentially with 
the correlation length 1. The correlation length may have influence on the fluctuations \cite{pruneau} and we plan to investigate this effect 
in the future. 

With this model we determine the central moments $\mu_i$ up to the fourth order and the volume-independent ratios of cumulants
$\chi_i$
\begin{eqnarray}
S\sigma & = & \frac{\chi_3}{\chi_2} = \frac{\mu_3}{\mu_2}\\
\kappa \sigma & = & \frac{\chi_4}{\chi_2} = \frac{\mu_4}{\mu_{2}} - 3 \mu_2 \\
\frac{\kappa \sigma^4}{\bar n} & = & \frac{\chi_4}{\chi_1} = \frac{\mu_4 - 3 \mu_2^2}{\mu_1}\,  ,
\end{eqnarray}
where we have used $\sigma$ for the width, $S$ for the skewness, and $\kappa$ for the kurtosis of the multiplicity distribution.


\section{A warm-up exercise: baryon number conservation and modifications}

Before we embark on more realistic simulation, let us test the model and present the mere effect of baryon number 
conservation and limited acceptance. 
We inspect a sample of $5.10^7$~events with fixed number of wounded nucleons $N_w = 338$ and other 
parameters set to reproduce Au+Au collisions at $\sqrt{s_{NN}} = 19.6$~GeV: $y_m = 1.019$ and 
$N_{B\bar B} = 16.946$.

\begin{figure}[t]
\centerline{\includegraphics[width=0.33\textwidth]{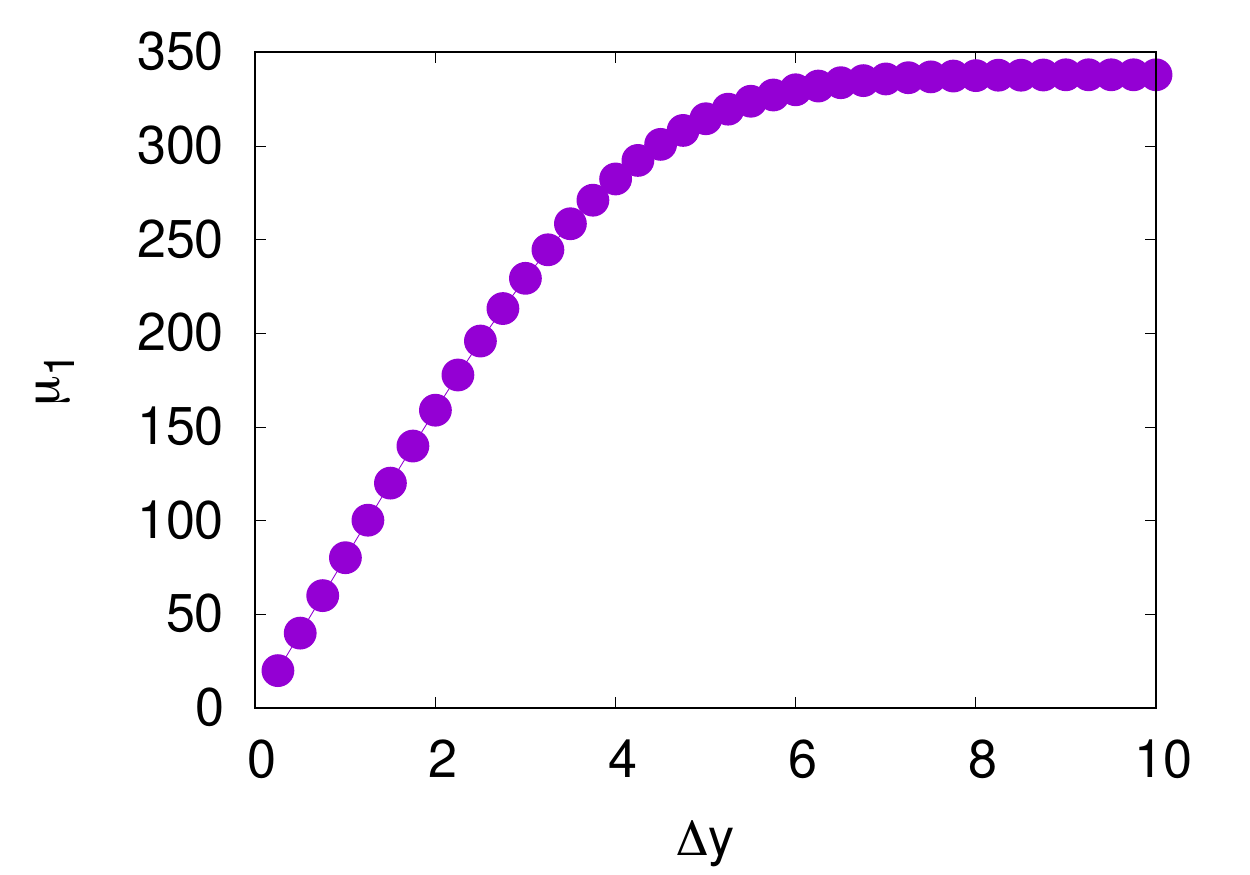}
\includegraphics[width=0.33\textwidth]{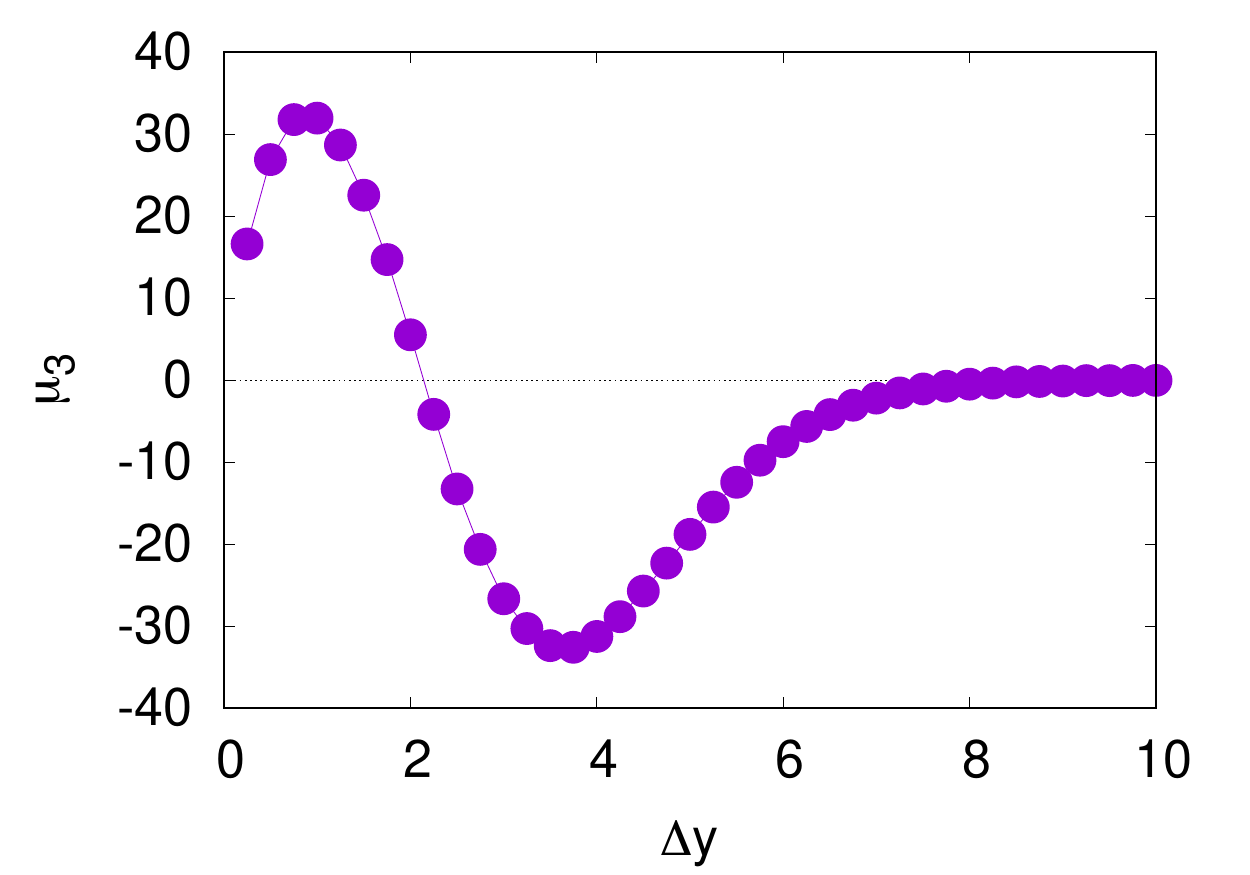}
\includegraphics[width=0.33\textwidth]{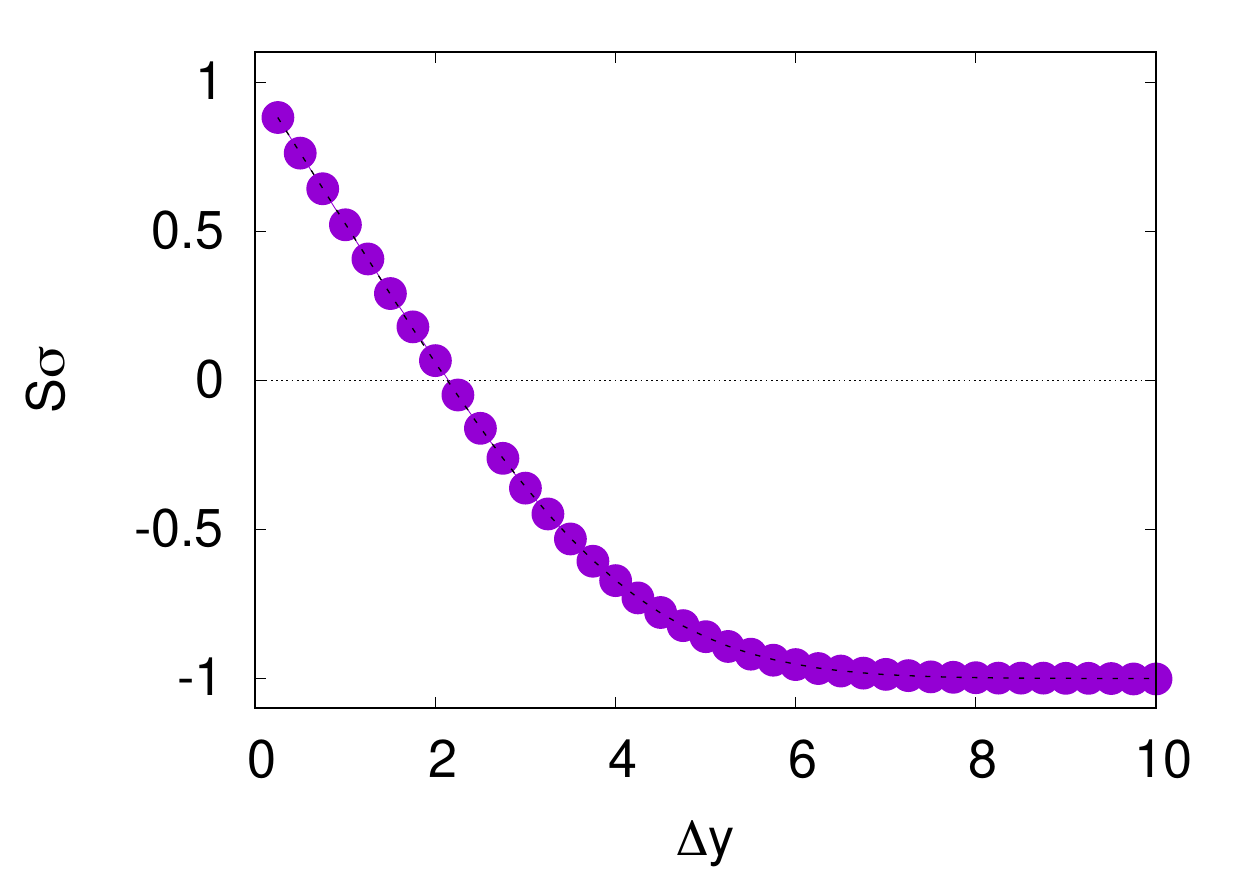}}
\centerline{\includegraphics[width=0.33\textwidth]{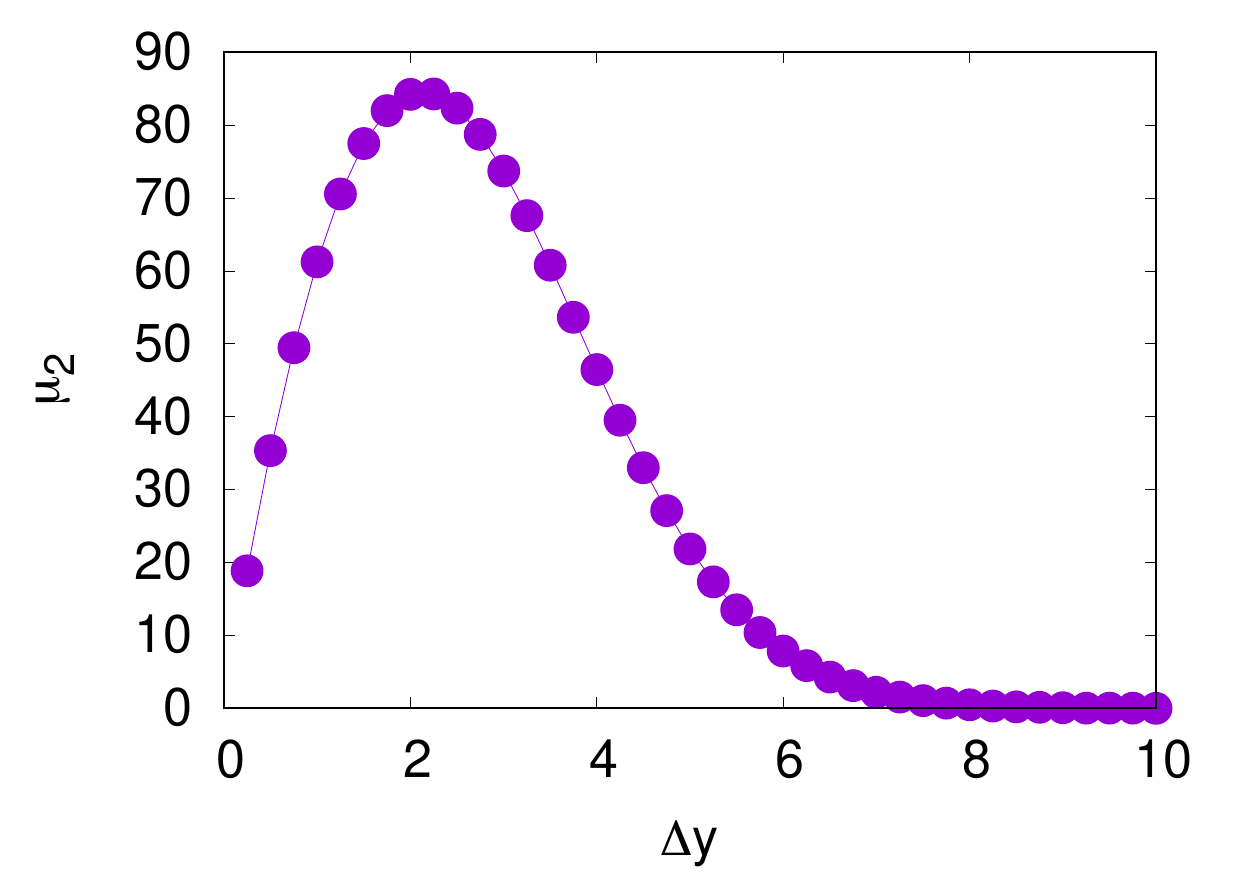}
\includegraphics[width=0.33\textwidth]{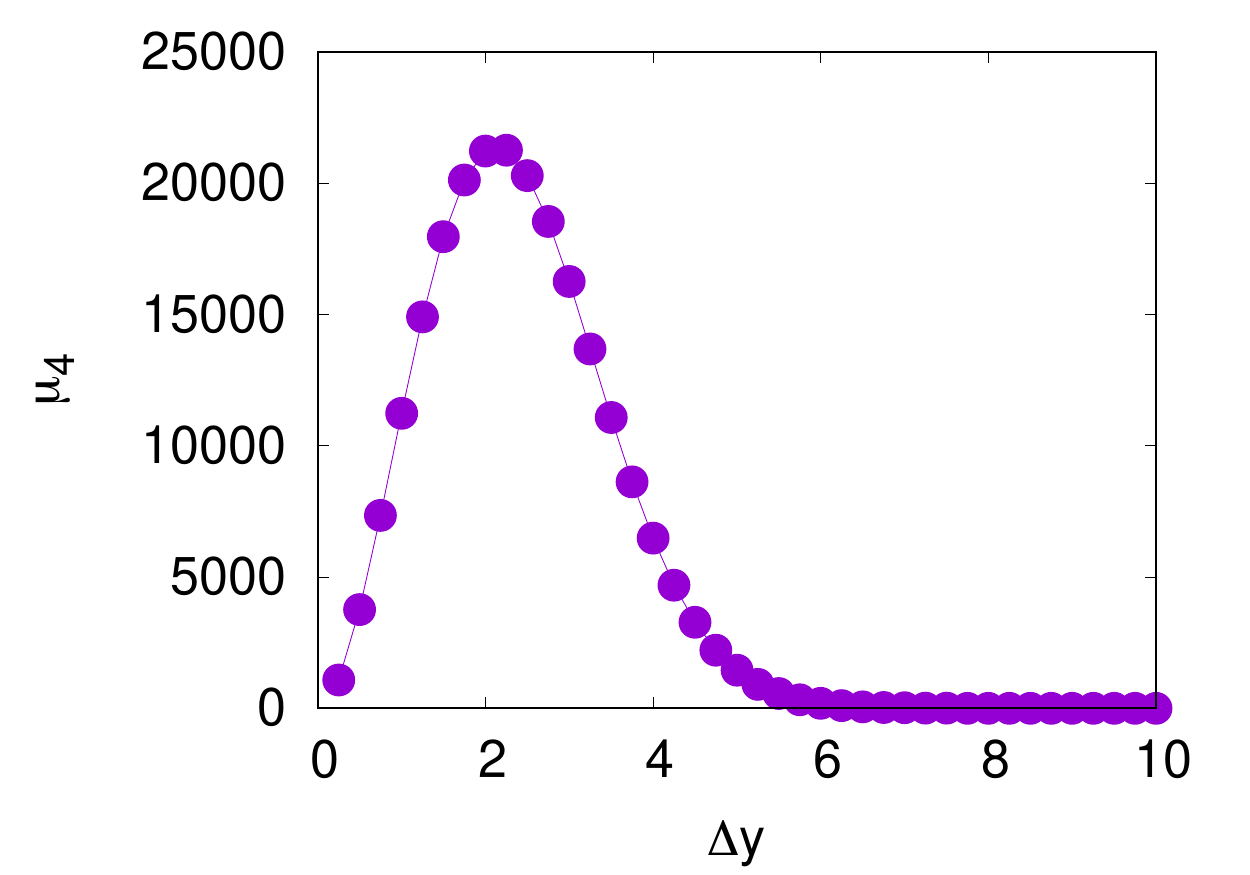}
\includegraphics[width=0.33\textwidth]{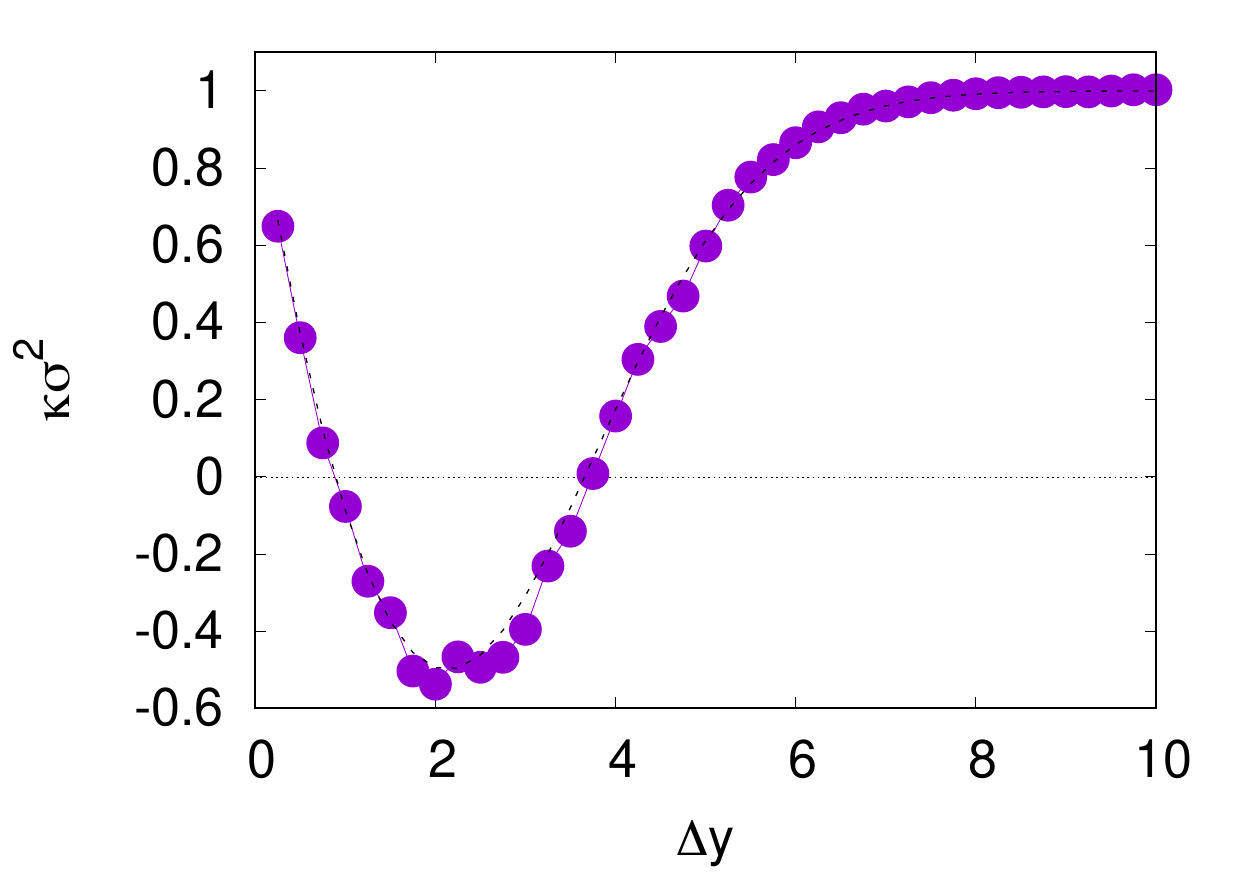}}
\caption{
Central moments of the baryon number distribution and the volume-independent  ratios $S\sigma$ and
$\kappa\sigma^2$ as functions of the rapidity window $\Delta y$. Parameter settings: $N_w = 338$, $y_m = 1.019$, and 
$N_{B\bar B} = 16.946$. The dotted curves in the two right panels show the dependences (\ref{e:binS}) and (\ref{e:binkappa}).
}
\label{f:Bnumbcon}
\end{figure}

In Figure~\ref{f:Bnumbcon} 
we plot the central moments and volume independent cumulant ratios of the net baryon number 
distribution, as functions of the width of the acceptance window $\Delta y$ around midrapidity. 
We show the central moments since they are the primary quantities which could be determined experimentally
and the other ratios are based on them. Thus one can relate the behaviour of the $\Delta y$ dependences
of the scaled skewness and the scaled kurtosis to the behaviour of the central moments. 
As $\Delta y$ increases, the mean value $\mu_1$ saturates at $N_w$ while the higher moments go to 
0 as there is no room for fluctuations due to baryon number conservation. The dependences are actually 
described by the  binomial distribution where the probability $p$ equals $\mu_1(\Delta y)/N_w$. This is 
demonstrated in the plots for the scaled skewness $S\sigma$ and the scaled kurtosis $\kappa\sigma^2$. The simulated points
are compared to the curves obtained for the binomial distribution
\begin{eqnarray}
S\sigma & = & 1 - 2p = 1 - 2\frac{\mu_1(\Delta y)}{N_w}   \label{e:binS}\\
\kappa \sigma^2 & = & 1 - 6p + 6 p^2 = 1 - 6 \frac{\mu_1(\Delta y)}{N_w} + 6\left ( \frac{\mu_1(\Delta y)}{N_w} \right )^2\,  .
\label{e:binkappa}
\end{eqnarray}
The observed dependences result just from the rapidity distribution of baryons and antibaryons 
combined with the binomial distribution of the number of accepted nucleons. 

The third central moment and consequently the skewness is  positive for small $\Delta y$ and then turns negative. This means, that 
in small rapidity windows numbers larger than the mean are rather populated. As the rapidity window gets larger, 
the total baryon number limits the possible observed numbers and they are more often populated below the mean.

Let us next look at how the results are modified, if now instead of baryons one measures the fluctuations of net protons. 
\begin{figure}[t]
\centerline{\includegraphics[width=0.33\textwidth]{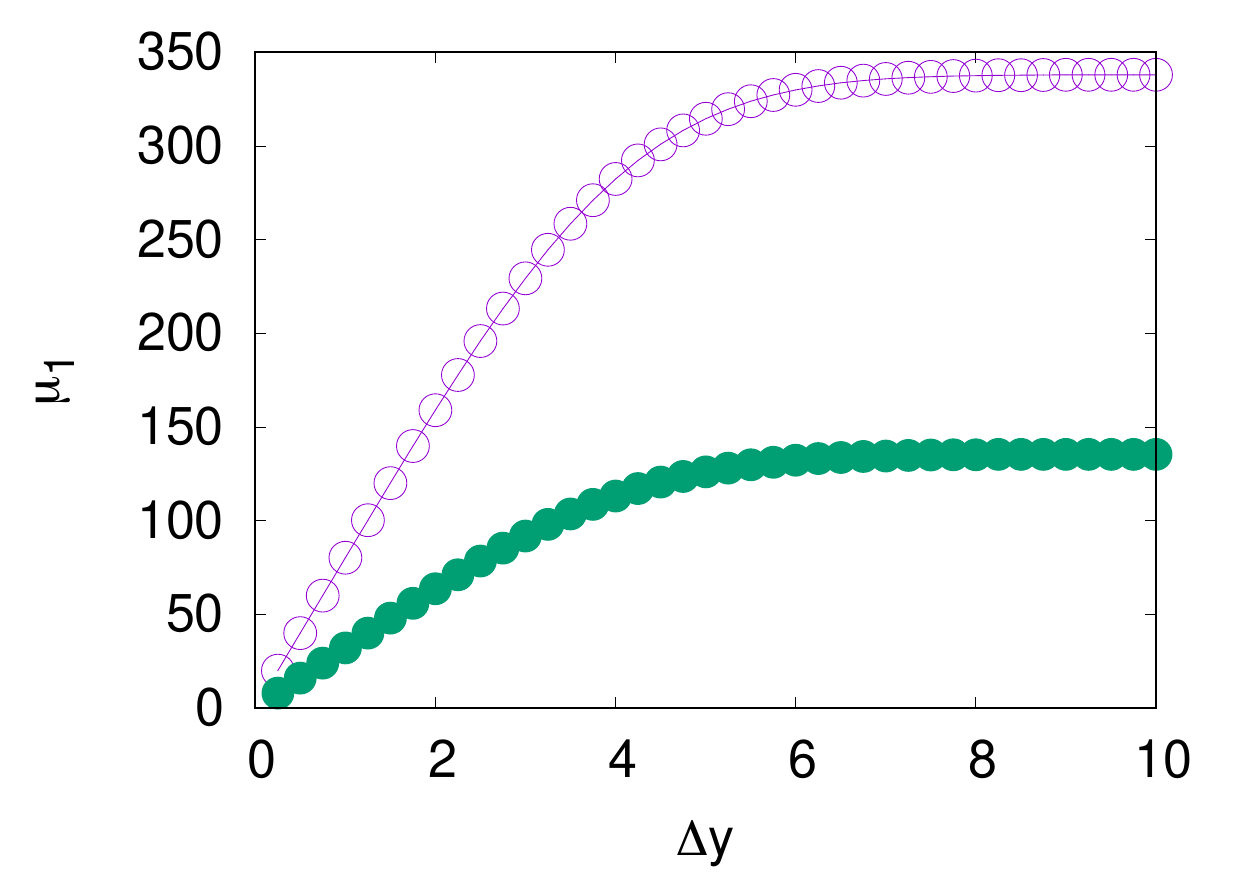}
\includegraphics[width=0.33\textwidth]{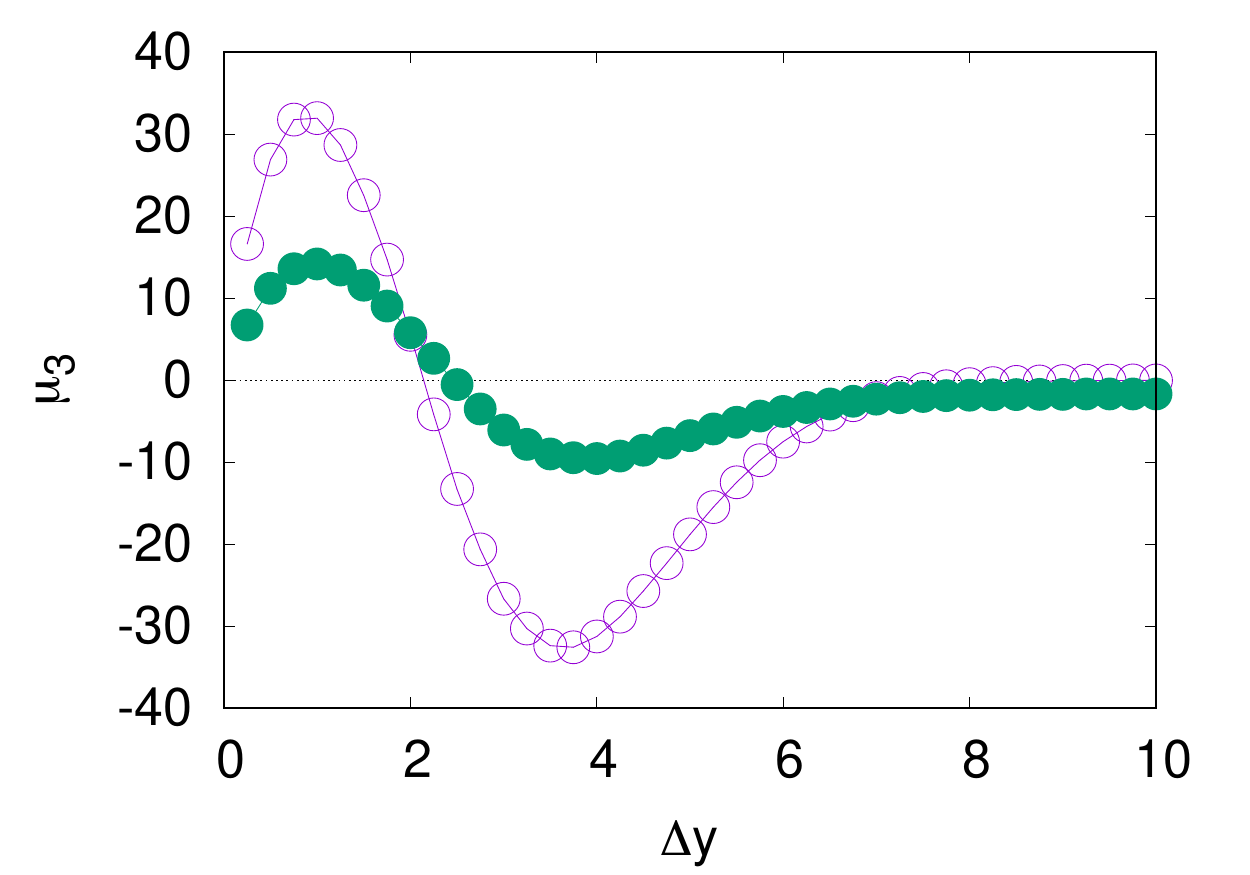}
\includegraphics[width=0.33\textwidth]{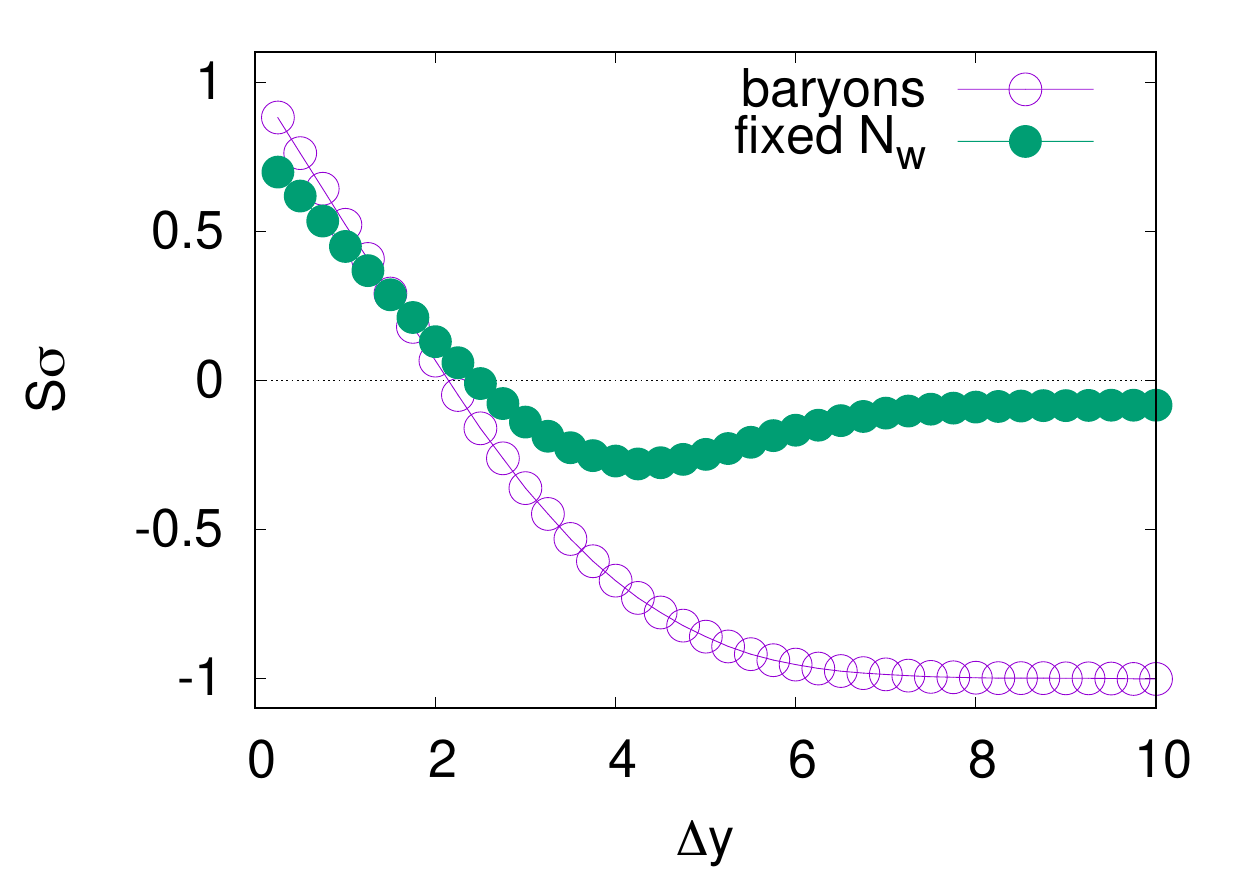}}
\centerline{\includegraphics[width=0.33\textwidth]{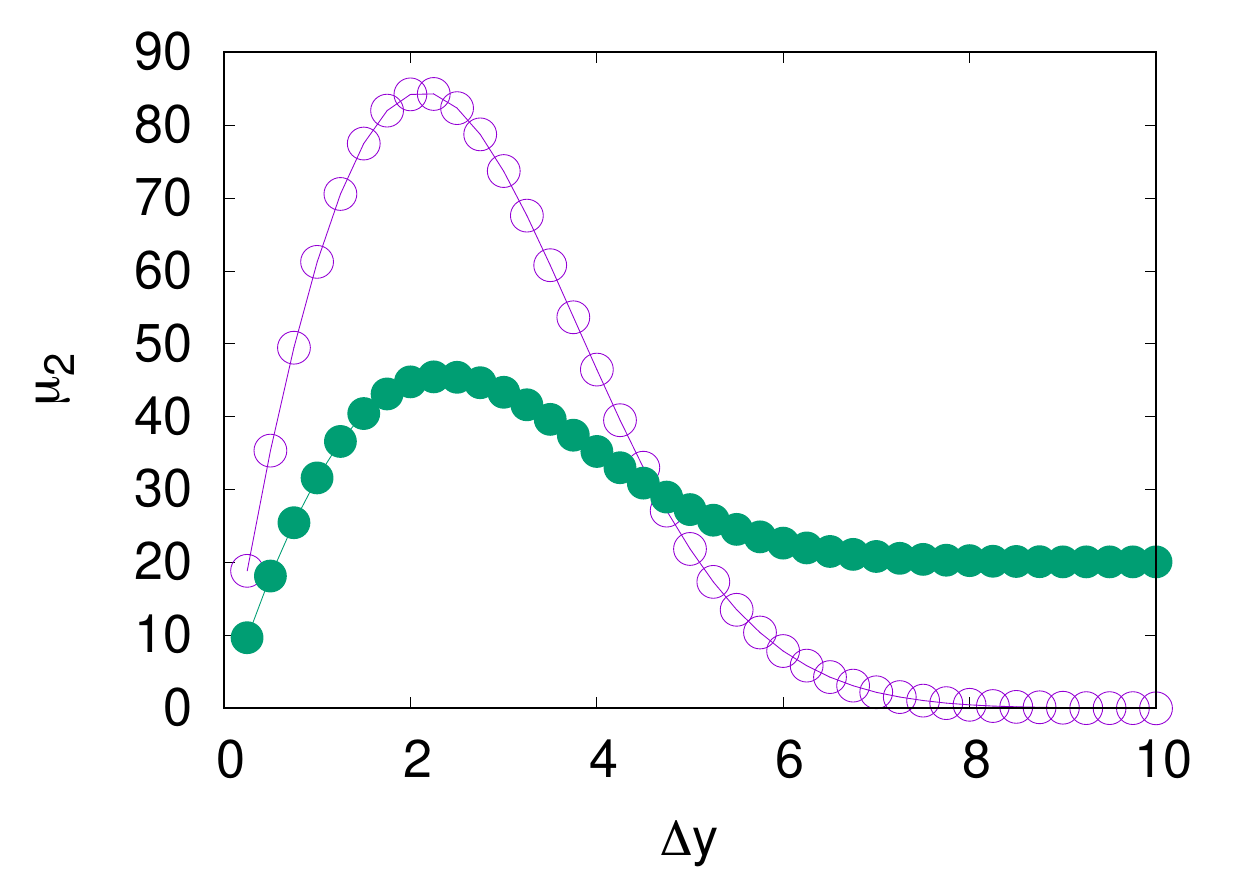}
\includegraphics[width=0.33\textwidth]{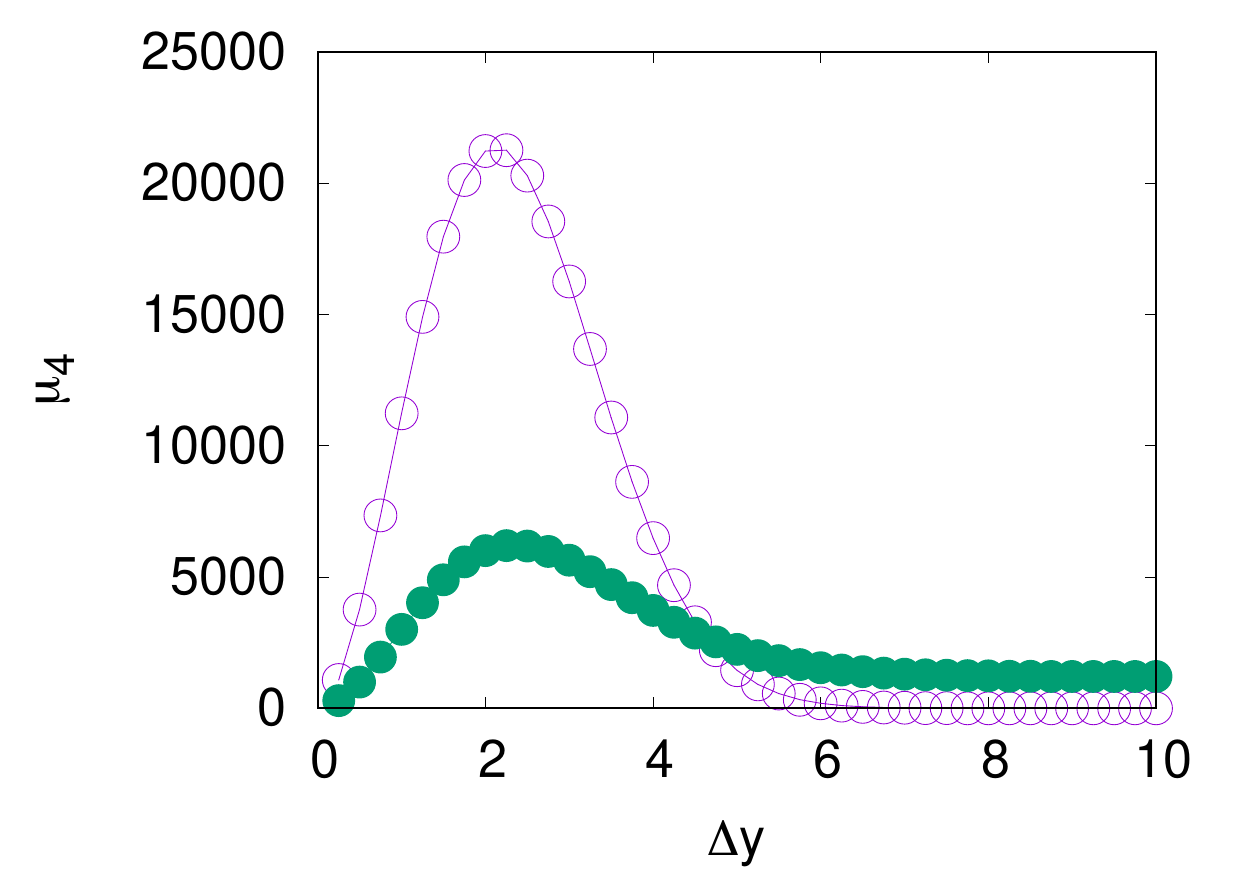}
\includegraphics[width=0.33\textwidth]{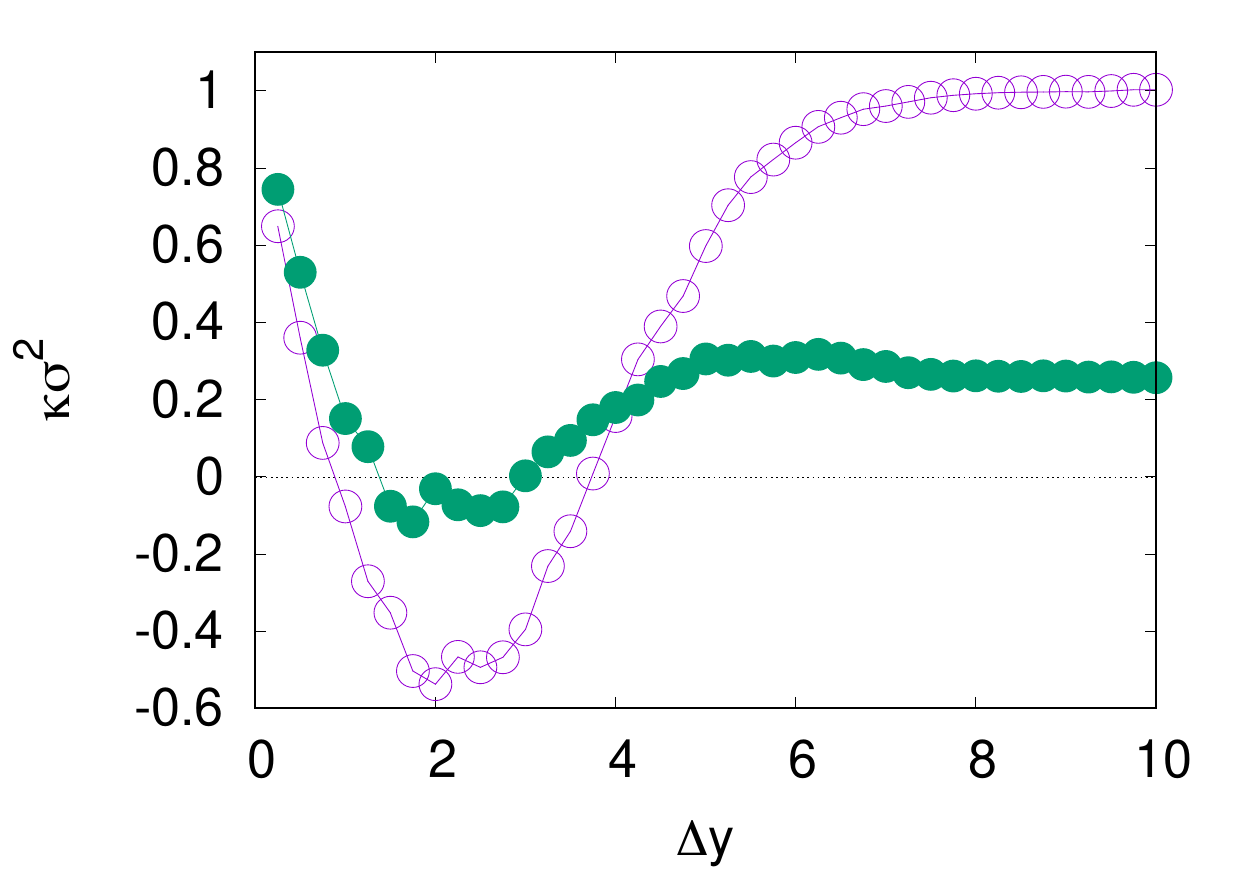}}
\caption{
Central moments of the net proton number distribution and the volume-independent  ratios $S\sigma$ and
$\kappa\sigma^2$ as functions of the rapidity window $\Delta y$. Parameter settings as in Fig.~\ref{f:Bnumbcon}.
Curves for the net proton number (full green circles) are compared with the curves for net baryons (open purple
circles) shown in Fig.~\ref{f:Bnumbcon}. The number of wounded nucleons is fixed in these simulations. 
}
\label{f:Pnumcon}
\end{figure}	
They are shown in Fig.~\ref{f:Pnumcon}.
The rough features of the baryon number distribution are inherited here. However, 
just by looking at the net protons---whose number is not conserved---there are qualitative differences  from 
the net baryons and the binomial distribution is no longer applicable. This is due to two reasons: firstly, the produced nucleon-antinucleon 
pairs populate protons and neutrons  with equal probability. Secondly, also the wounded nucleons may change isospin. The result 
is that now the total number of net protons is not fixed and fluctuates from event to event. This is seen in the behaviour 
of the higher central moments which saturate at a non-vanishing values for large rapidity windows. The asymptotic values of 
$S\sigma$ and $\kappa\sigma^2$ are modified, as well.

\begin{figure}[t]
\centerline{\includegraphics[width=0.33\textwidth]{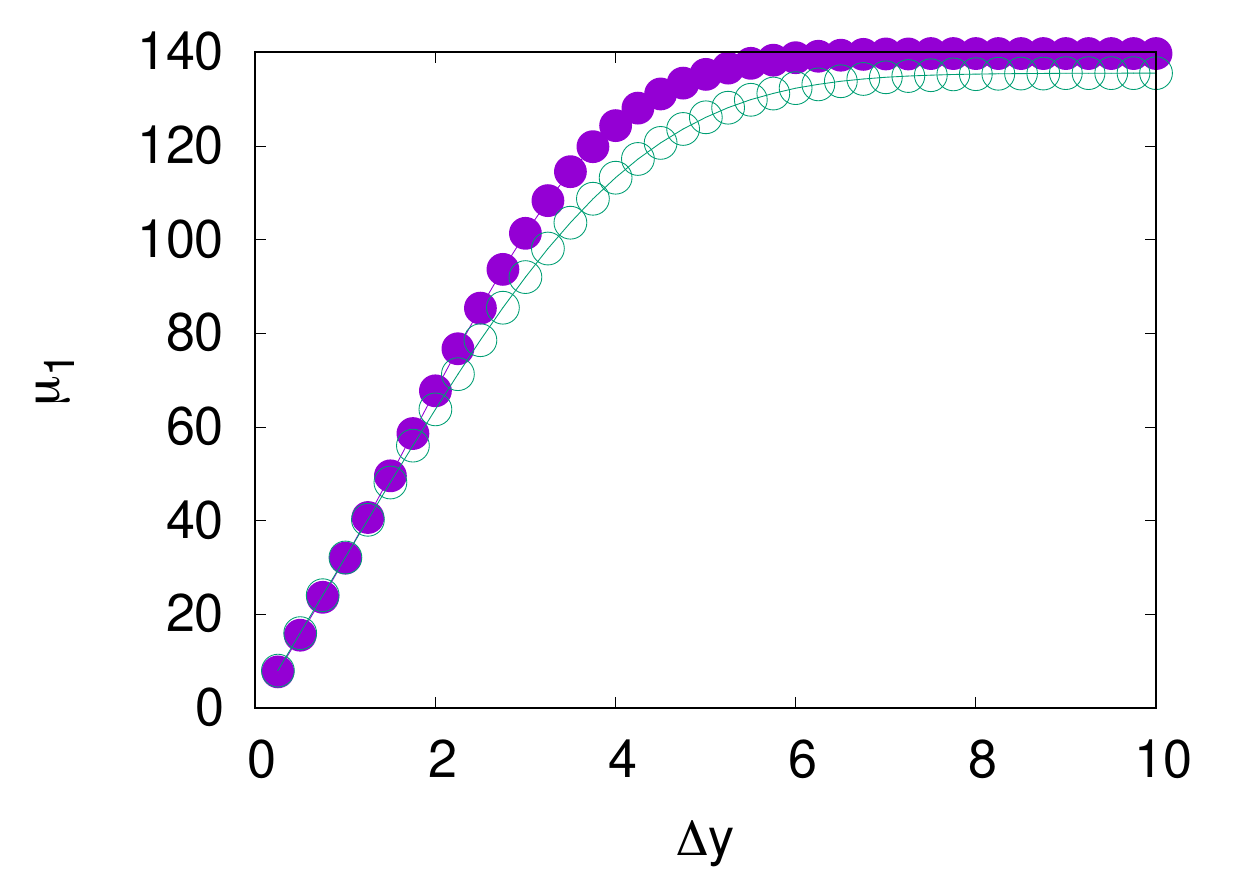}
\includegraphics[width=0.33\textwidth]{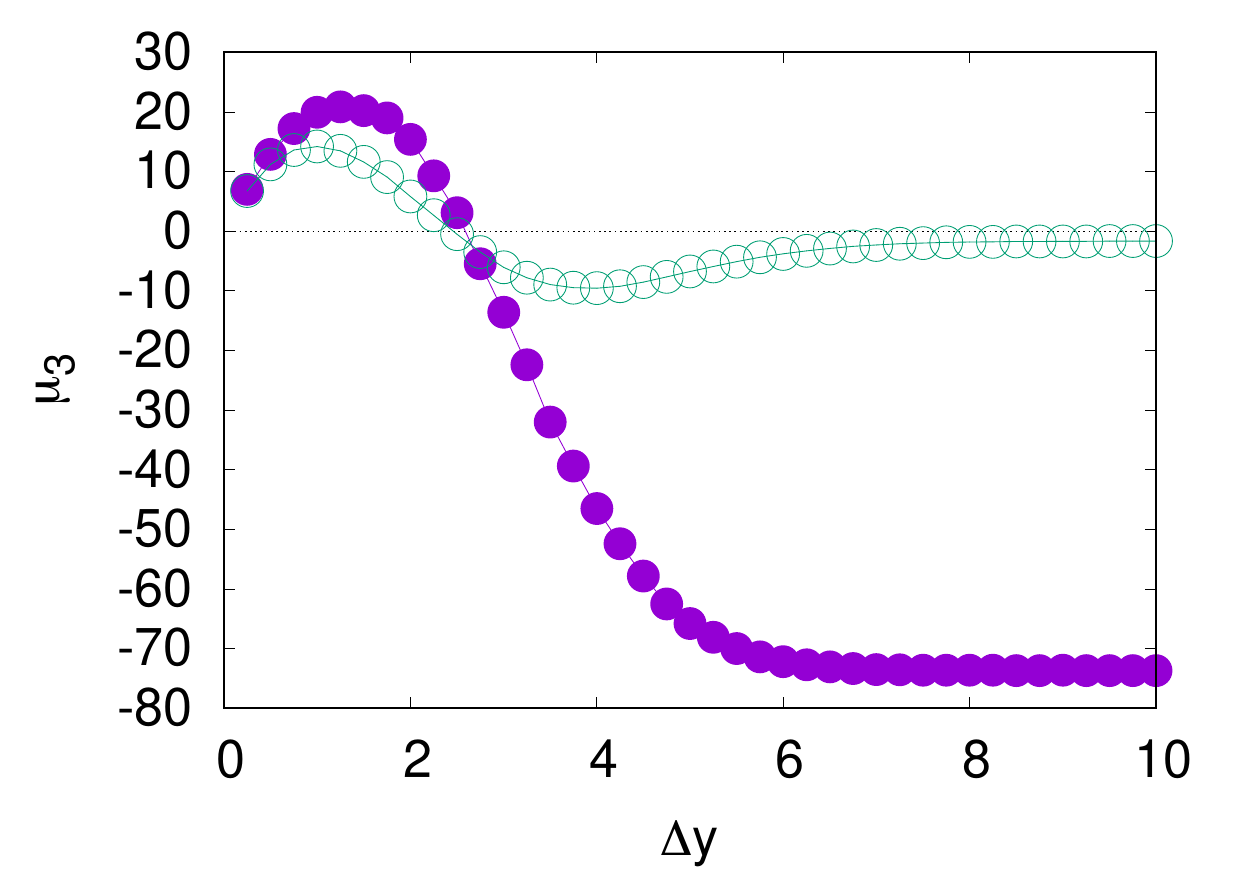}
\includegraphics[width=0.33\textwidth]{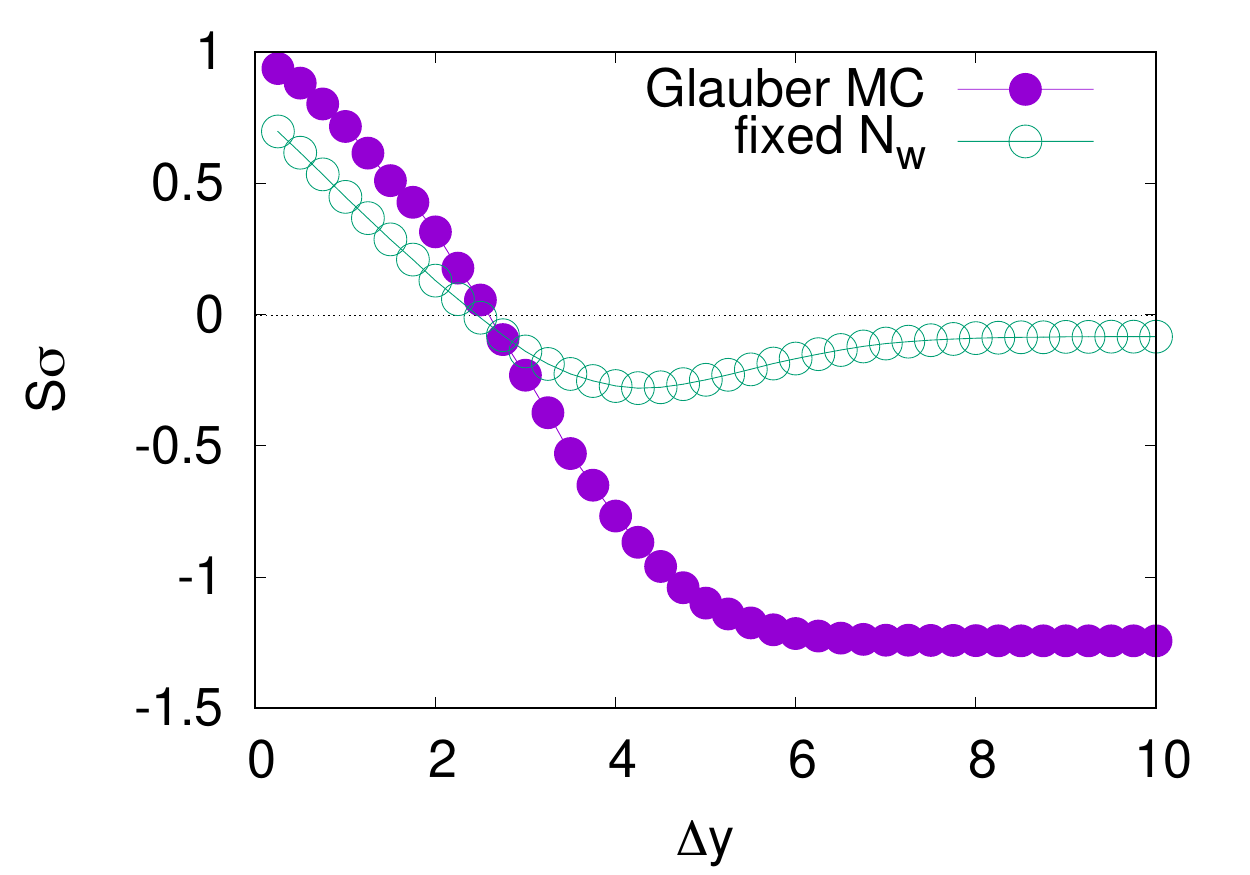}}
\centerline{\includegraphics[width=0.33\textwidth]{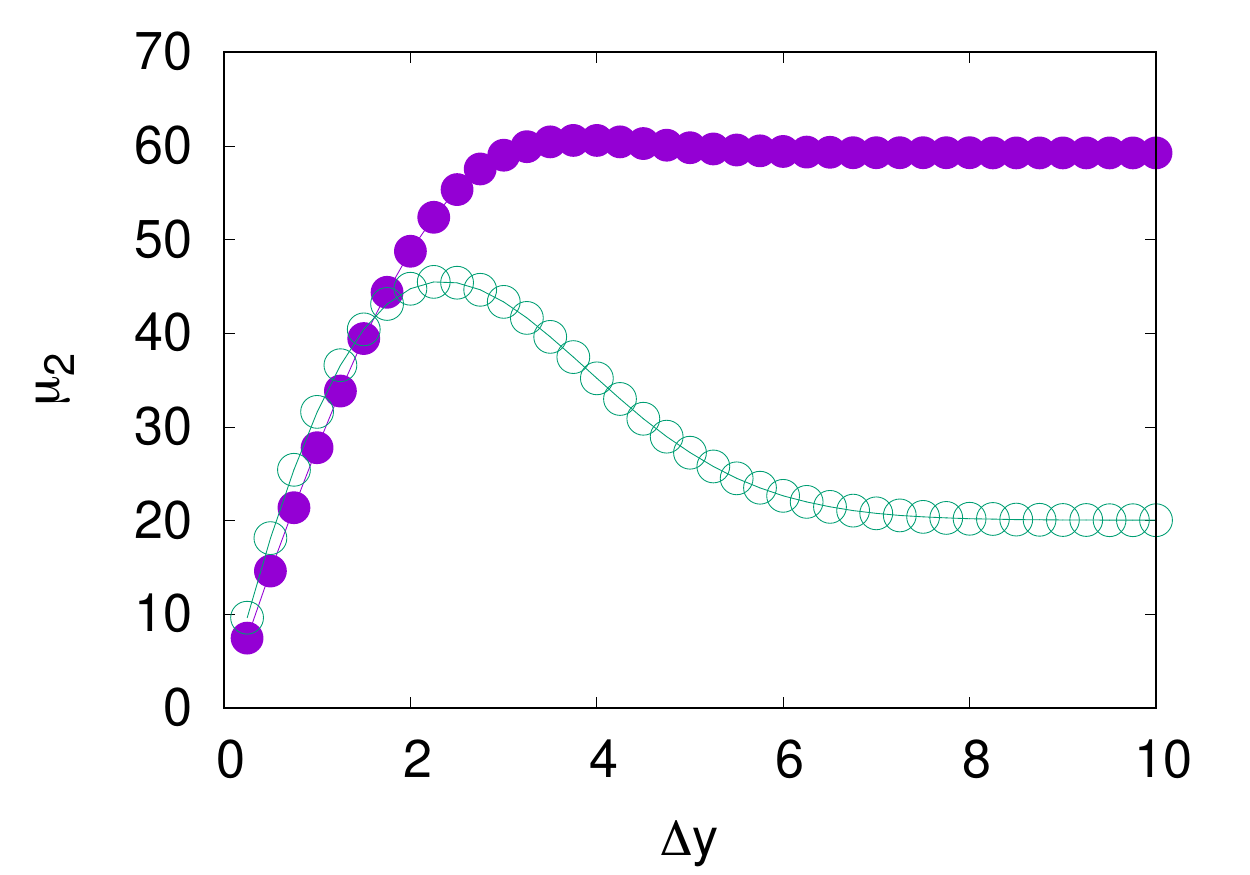}
\includegraphics[width=0.33\textwidth]{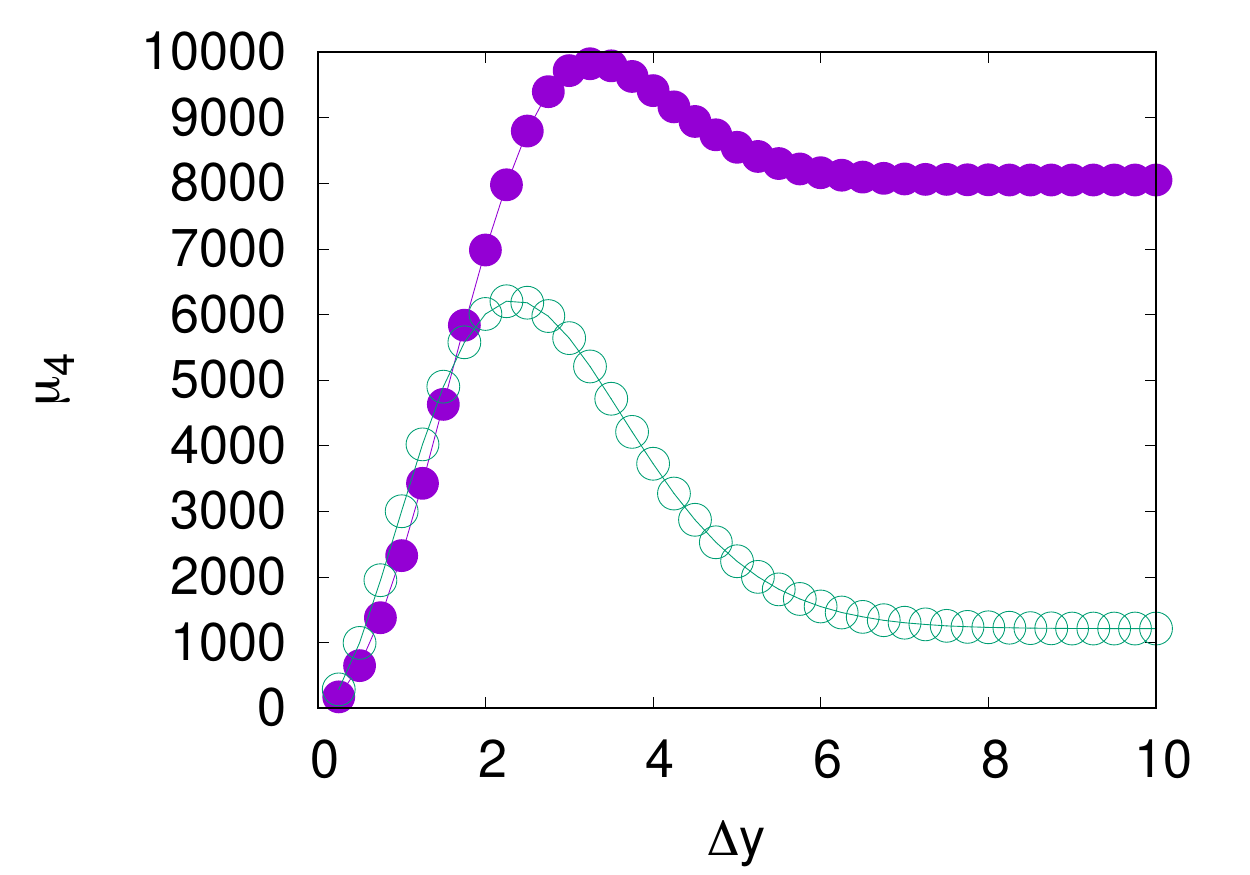}
\includegraphics[width=0.33\textwidth]{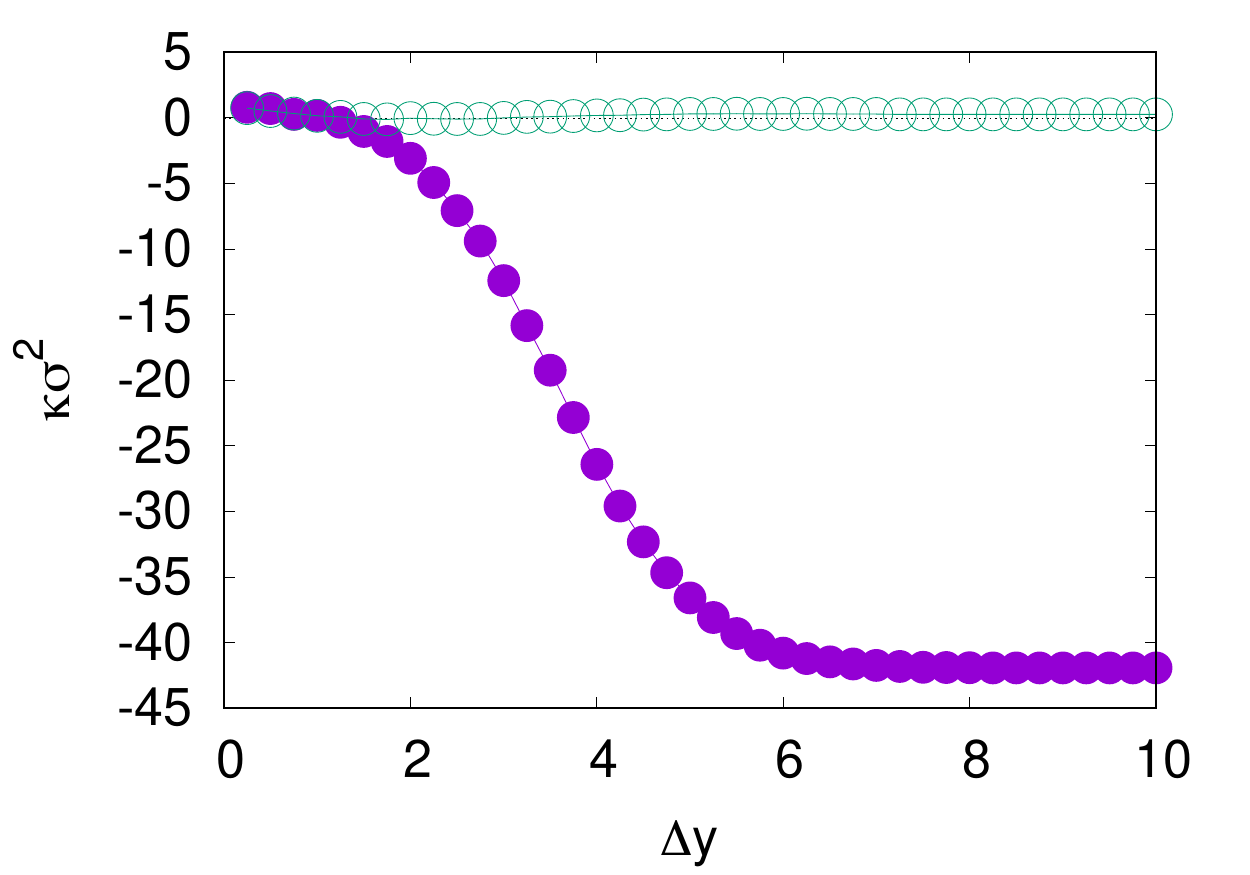}}
\caption{
Central moments of the net proton number distribution and the volume-independent  ratios $S\sigma$ and
$\kappa\sigma^2$ as functions of the rapidity window $\Delta y$. Parameter settings as in Figs.~\ref{f:Bnumbcon}
and \ref{f:Pnumcon}. Open green  symbols show the dependences for $N_w$ fixed to 338 (same curves as in Fig.~\ref{f:Pnumcon})
and full purple symbols show the dependences for fluctuating $N_w$ within the 0--5\% centrality class as determined from 
$1.2\times 10^6$ events generated by GLISSANDO.
}
\label{f:Nwfluct}
\end{figure}	
The next complication comes from the fluctuation of the number of participating nucleons. Centrality classes in an experiment 
are usually determined from the produced multiplicity. That usually scales with the relative deposited strength as defined in 
eq.~(\ref{e:rds}) which can be evaluated in GLISSANDO. 
Within the 0--5\% percentile with highest RDS among all events, which corresponds to the experimental class of 
0--5\% centrality, the number of wounded nucleons fluctuates. Figure~\ref{f:Nwfluct} shows how this influences the net-proton 
number fluctuations. In general, all central moments characterising fluctuations grow considerably with respect to the 
simulation with fixed $N_w$. 
Nevertheless, some qualitative features of their dependence on $\Delta y$ remain, e.g., the third moment flips the sign 
at the same $\Delta y$ as it did for fixed $N_w$.
These features are then reflected also in $S\sigma$ and $\kappa\sigma^2$.


\section{Results}

\begin{figure}[t]
\centerline{\includegraphics[width=0.33\textwidth]{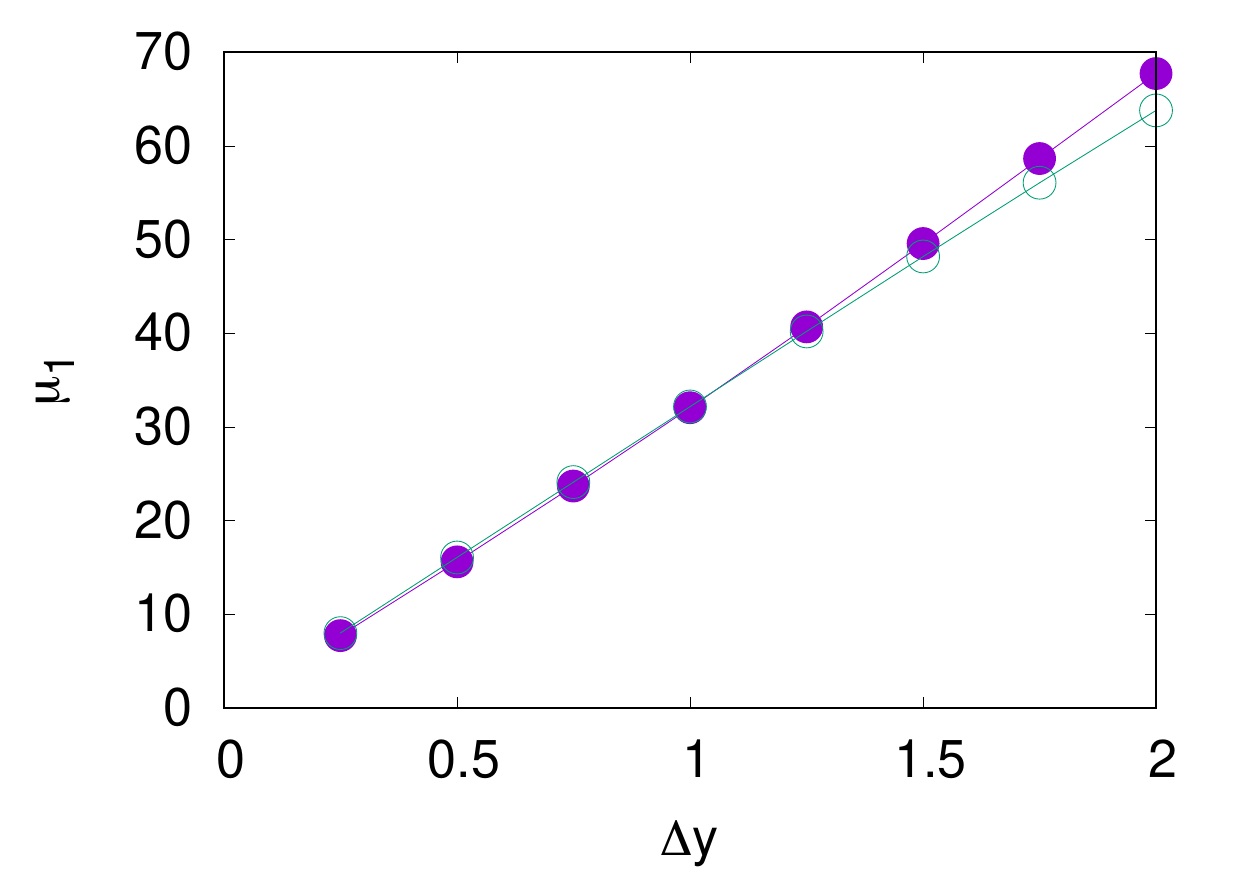}
\includegraphics[width=0.33\textwidth]{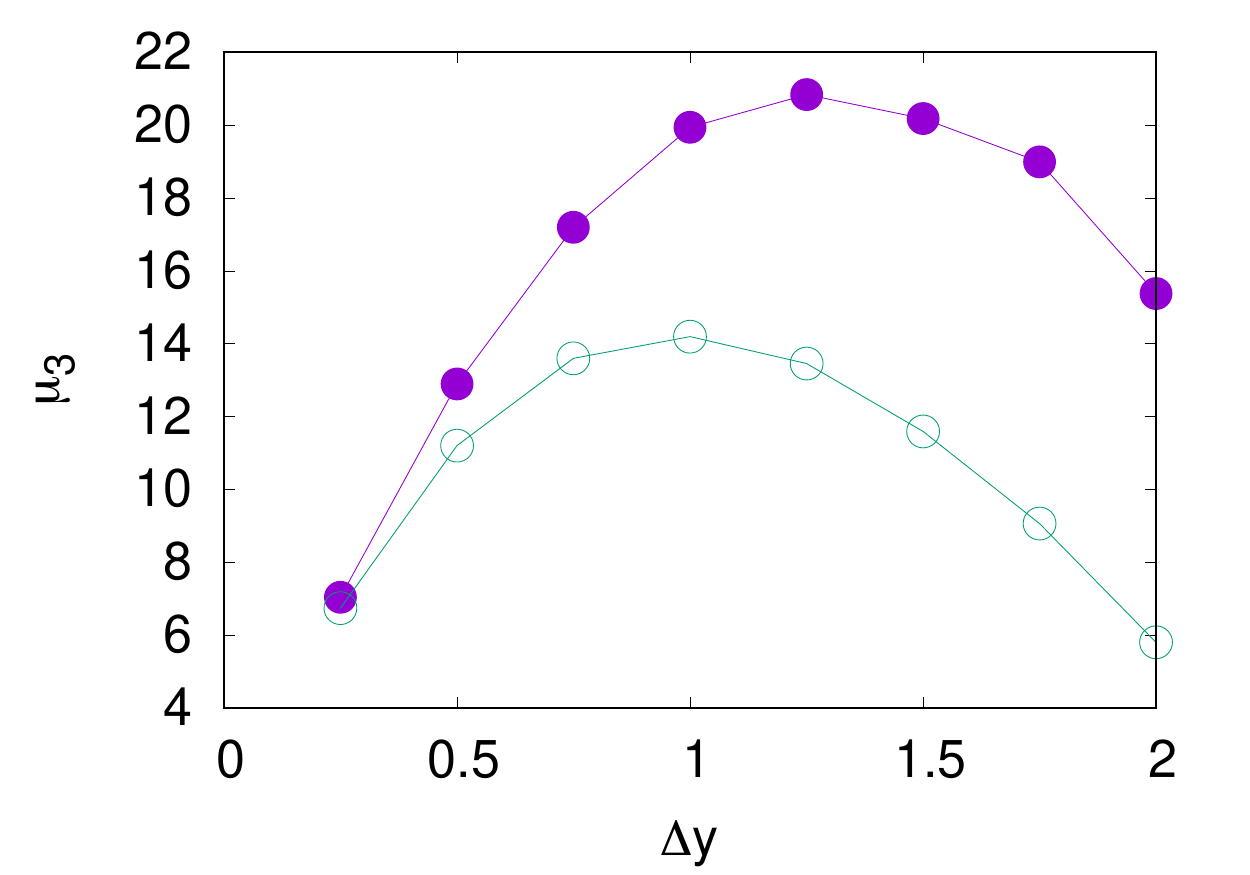}
\includegraphics[width=0.33\textwidth]{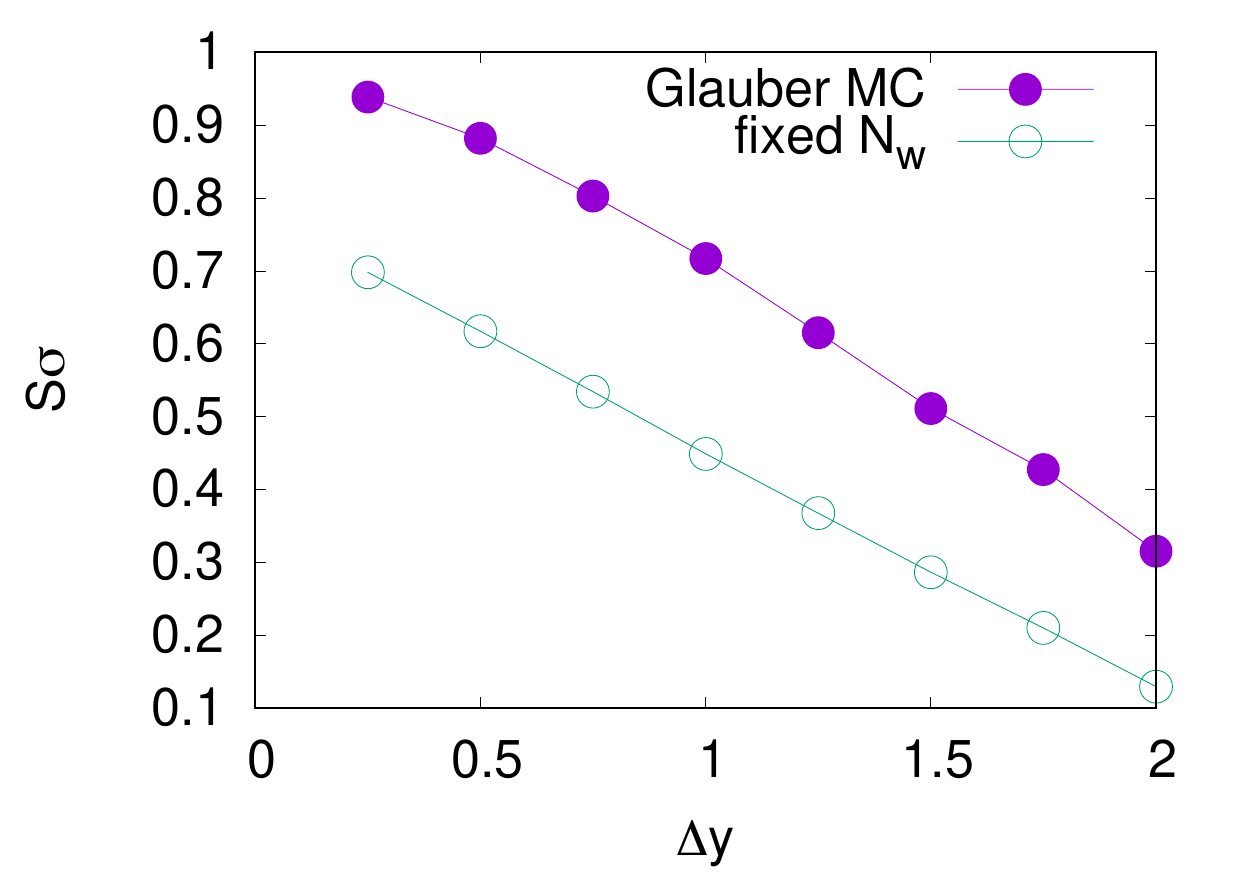}}
\centerline{\includegraphics[width=0.33\textwidth]{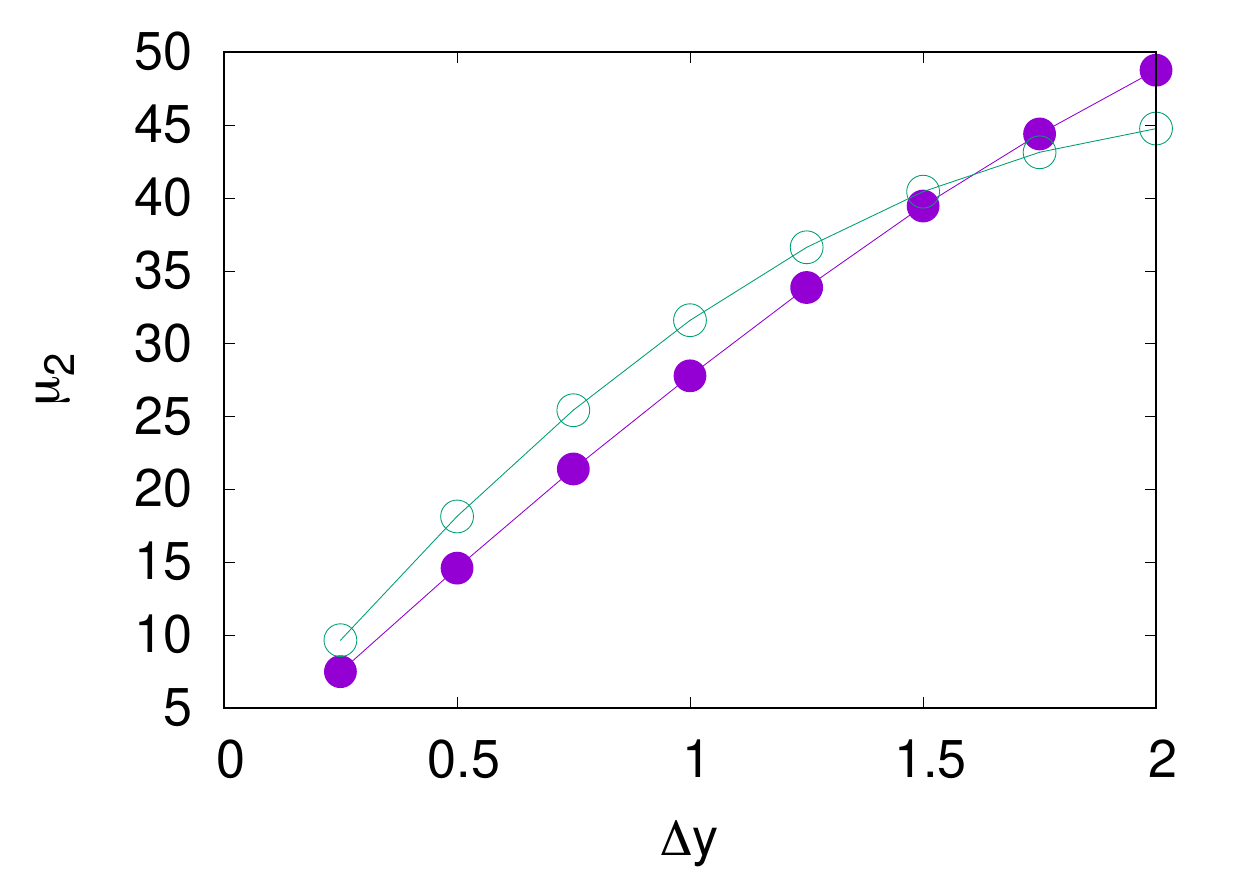}
\includegraphics[width=0.33\textwidth]{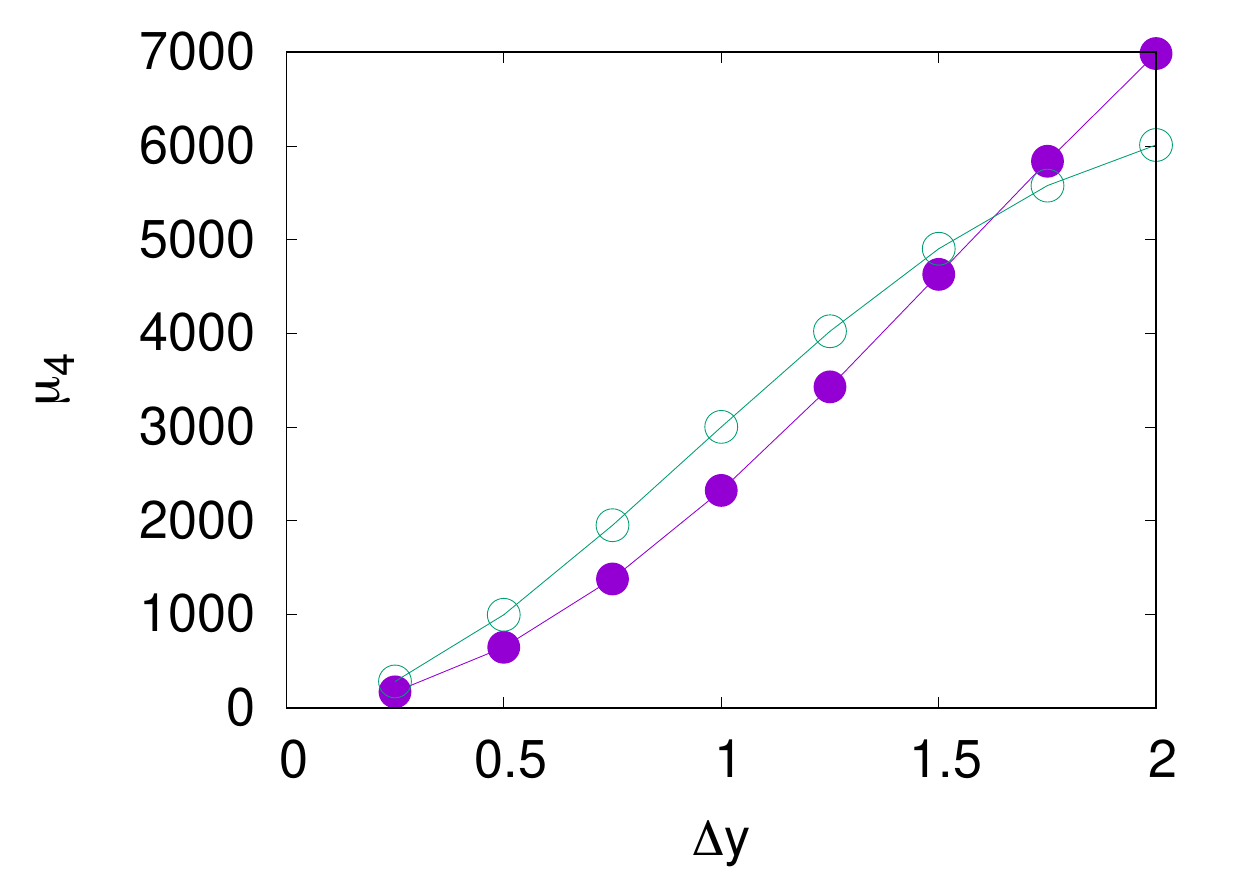}
\includegraphics[width=0.33\textwidth]{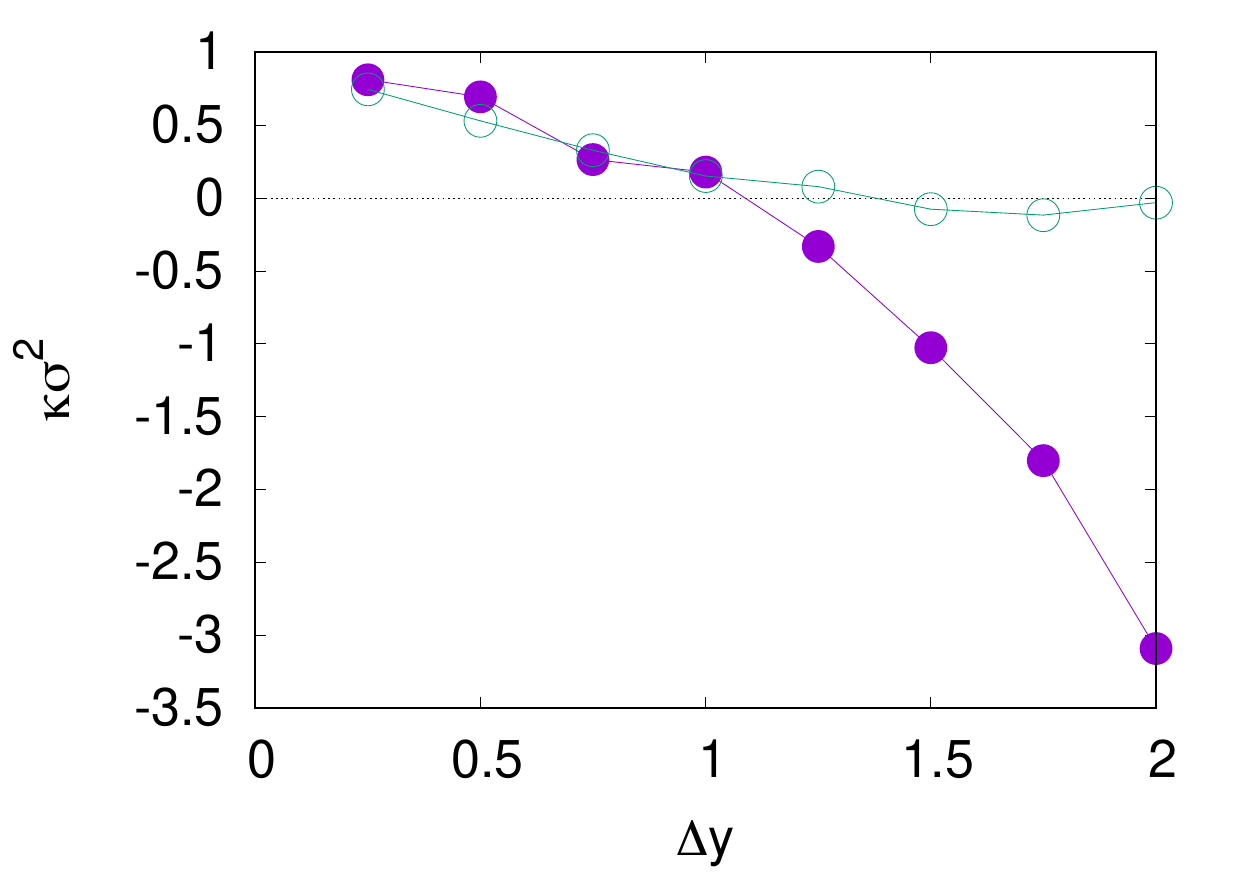}}
\caption{
Same as Figure~\ref{f:Nwfluct} but zoomed into smaller interval of $\Delta y$. 
}
\label{f:Nwzoom}
\end{figure}

Figures \ref{f:Bnumbcon}-\ref{f:Nwfluct} demonstrate the influence of the three basic effects on the observed 
fluctuations: baryon number conservation (Fig.~\ref{f:Bnumbcon}), measurement of only protons (Fig.~\ref{f:Pnumcon}), 
and fluctuations of the interacting volume (Fig.~\ref{f:Nwfluct}). From Fig.~\ref{f:Nwfluct} one would conclude that the latter 
influence is tremendous. However, no real detector in the current experiments covers 10 units of rapidity. Therefore, 
in Figure~\ref{f:Nwzoom} we zoom into a narrower $\Delta y$-interval of 2 units, which may realistically be covered
by the central detector, e.g. in the STAR experiment at RHIC. 
Here, the only large difference  among the central moments shows up for the third moment: results from Glauber MC simulations
lead to larger $\mu_3$.
This is then translated in about a constant increase of $S\sigma$ by 0.25 in the whole interval of $\Delta y$. Up to 
$\Delta y = 1$, the values of $\kappa \sigma^2$ for fluctuating and for fixed $N_w$ are practically identical. 
However, they start to depart from each other strongly beyond this limit. 

\begin{figure}[t]
\centerline{\includegraphics[width=0.33\textwidth]{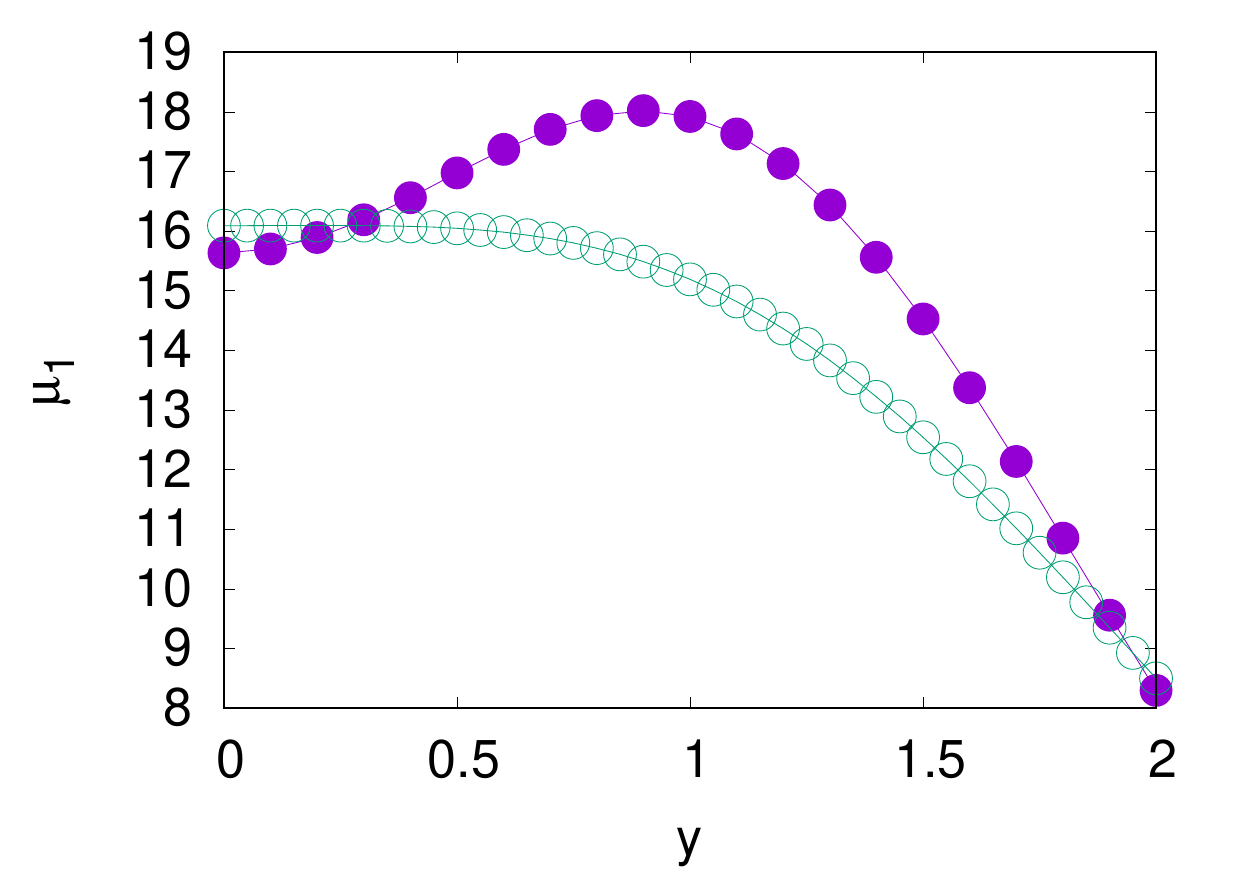}
\includegraphics[width=0.33\textwidth]{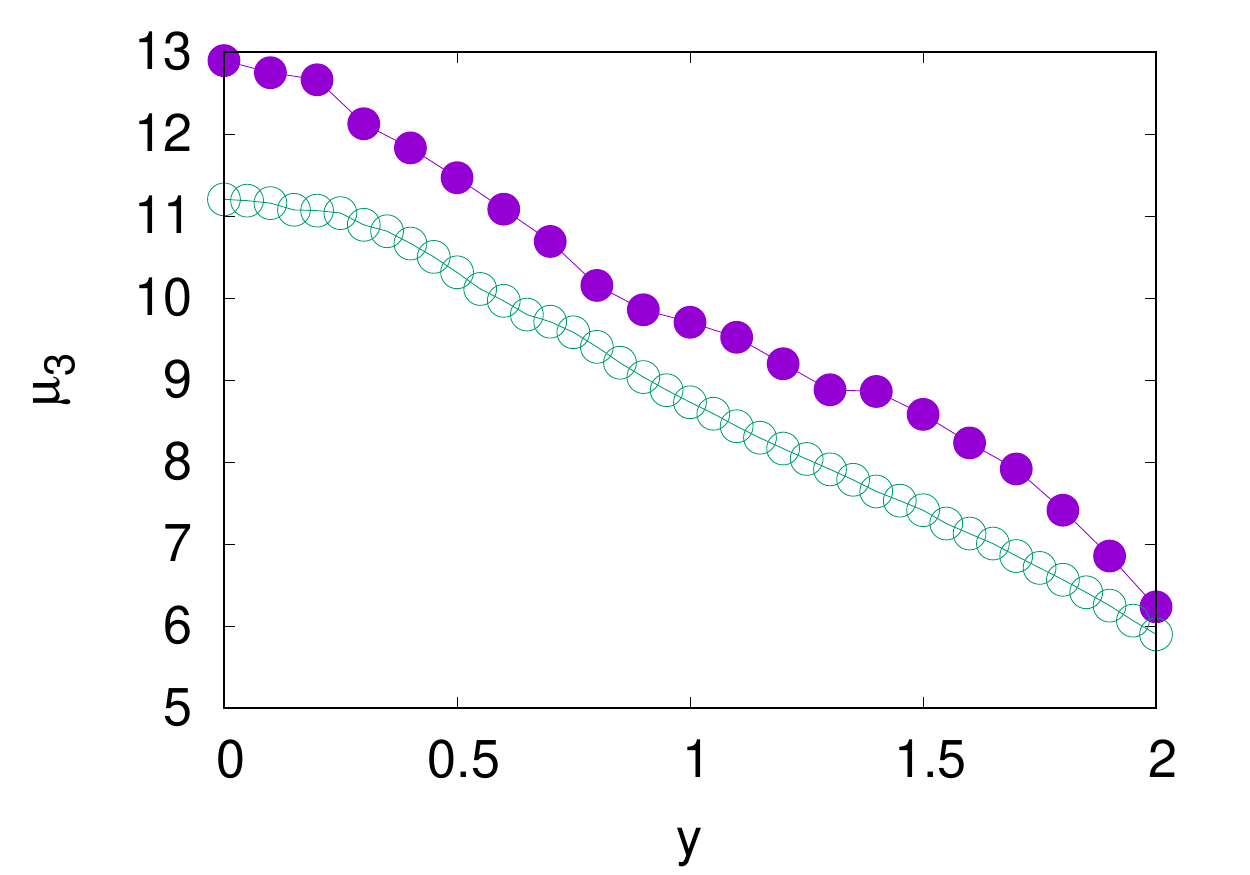}
\includegraphics[width=0.33\textwidth]{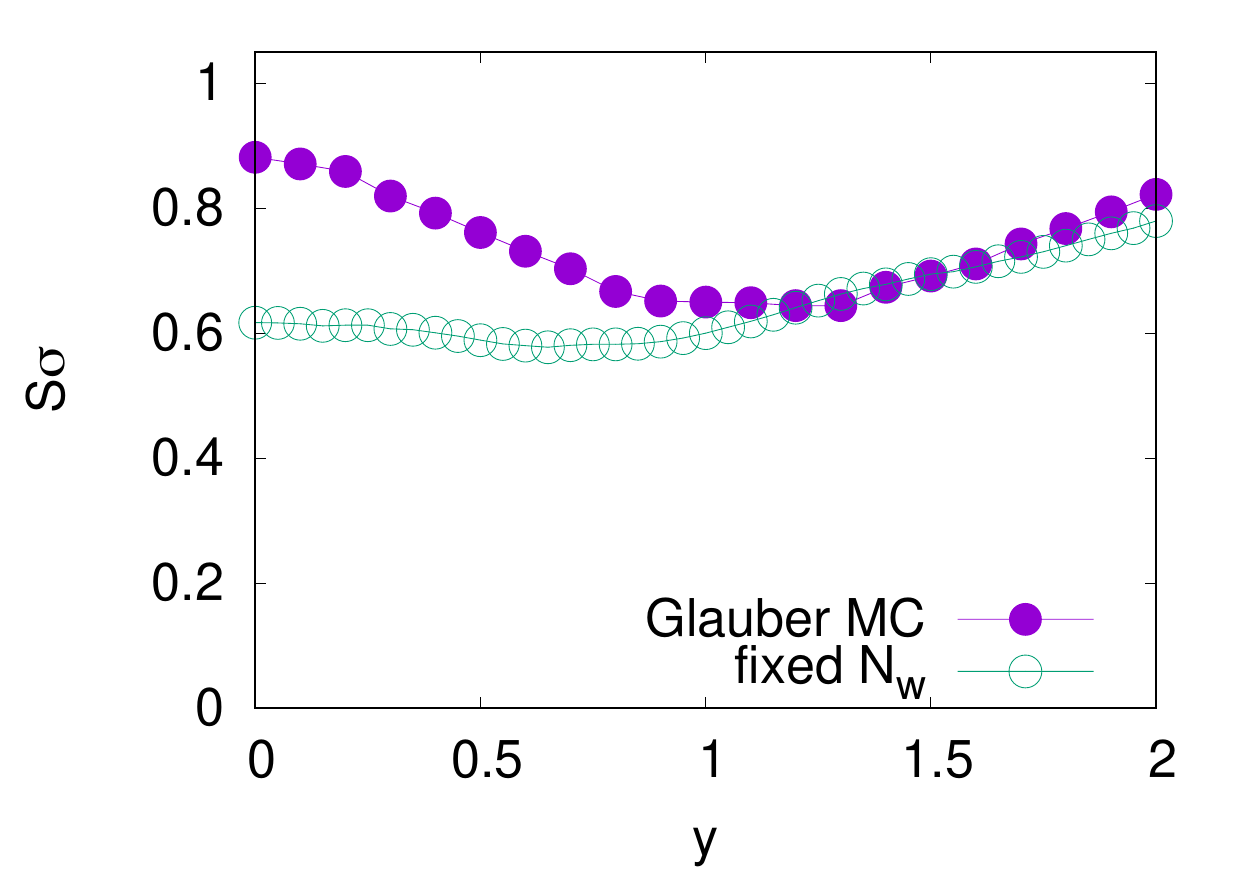}}
\centerline{\includegraphics[width=0.33\textwidth]{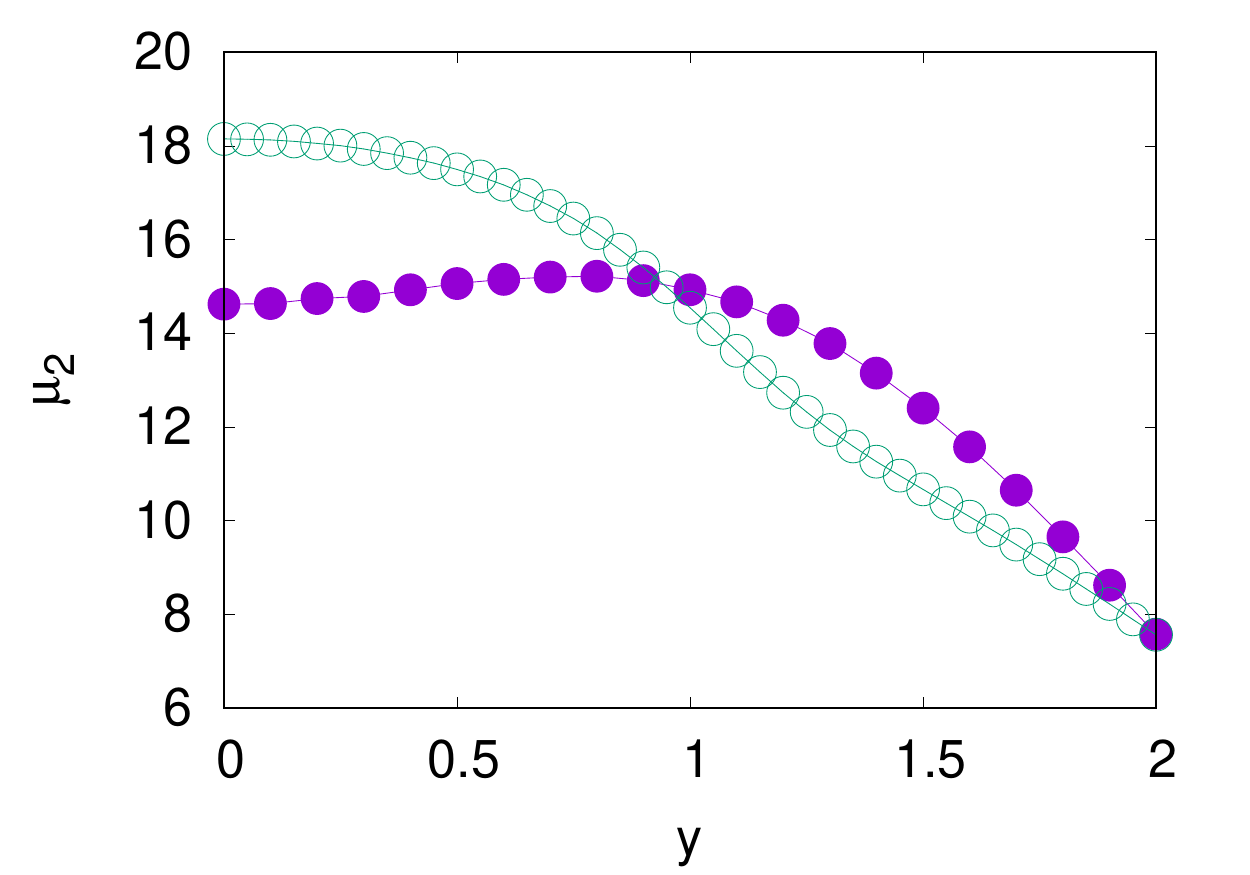}
\includegraphics[width=0.33\textwidth]{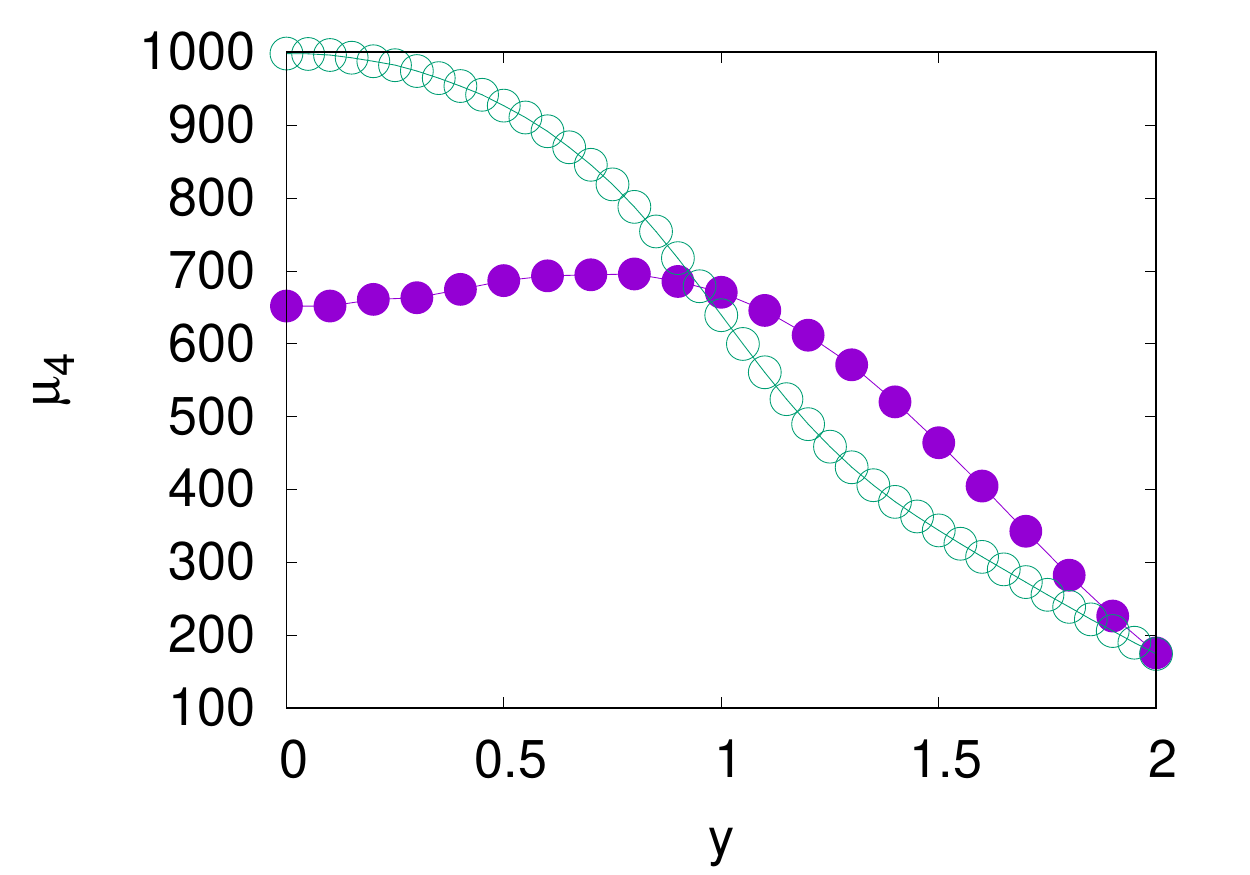}
\includegraphics[width=0.33\textwidth]{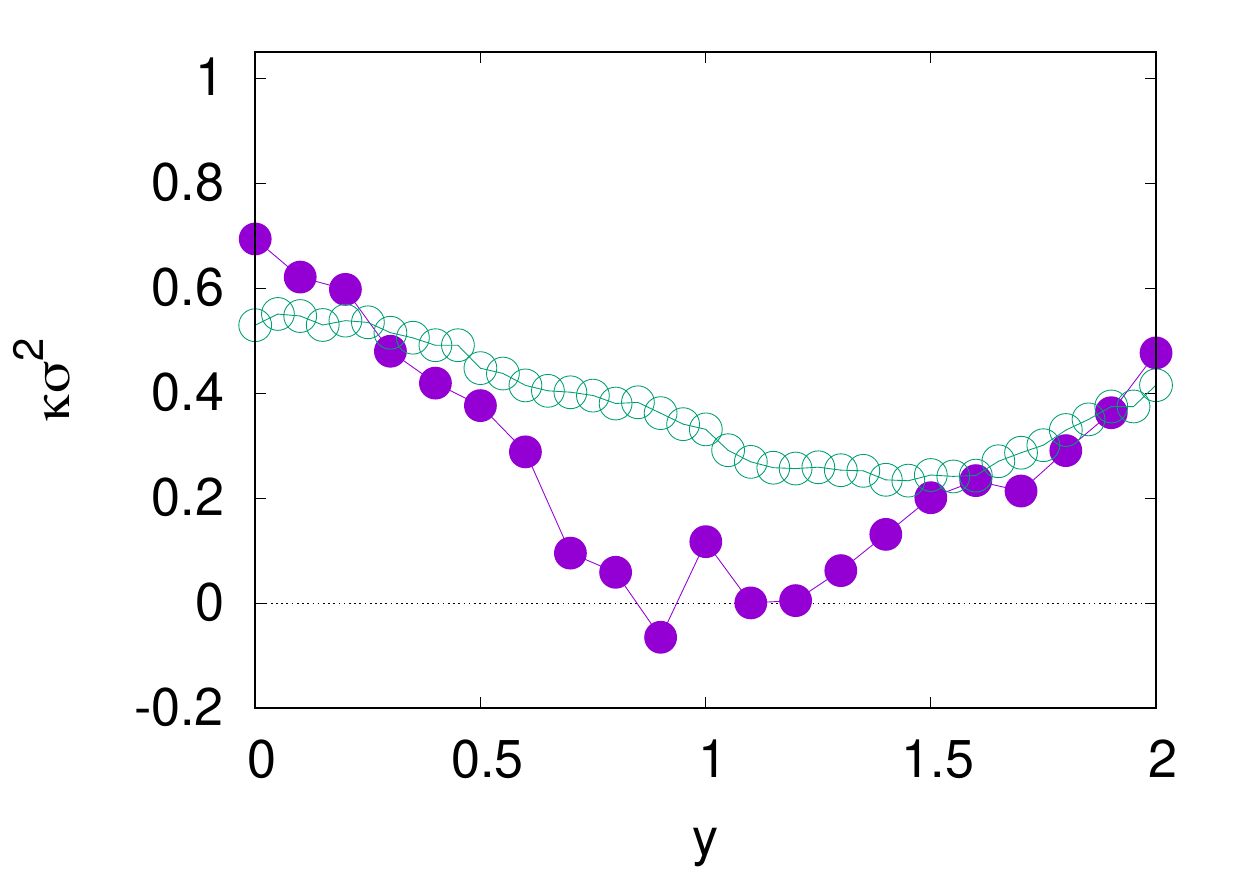}}
\caption{
Rapidity dependence of the central moments and volume-independent ratios $S\sigma$ and $\kappa\sigma^2$ of the net-proton distribution.
Parameter settings: $N_w = 338$, $y_m = 1.019$, and 
$N_{B\bar B} = 16.946$ (like in Fig.~\ref{f:Bnumbcon}).  
Open green symbols show results for fixed $N_w = 338$, full purple circles show results from $1.2\times 10^6$ of 0--5\% central 
Glauber MC events
generated by GLISSANDO. 
}
\label{f:Ydep}
\end{figure}

An interesting suggestion \cite{Brewer} has been presented also at this conference \cite{jasmine} that the critical point might be sought 
by inspecting the net-proton number fluctuations at different rapidities as an alternative to different collision energies. 
Our model includes the feature, that rapidity distributions of protons and antiprotons are different. Hence, we can make 
predictions for the rapidity dependence of the net-proton number fluctuations in absence of any critical behaviour.  
This is shown in Figure~\ref{f:Ydep}.
Simulations were done for a set of parameters that correspond to $\sqrt{s_{NN}} = 19.6$~GeV. Results have been extracted 
for a rapidity window of $\Delta y = 0.5$, which is put in different positions along the rapidity distribution. Figure~\ref{f:Ydep}
shows the dependences on the central value of the rapidity window. We compare the results obtained for fixed $N_w = 338$ 
($2 \times10^7$ events in the sample) with events from 0--5\% central collisions according to GLISSANDO Glauber MC ($1.2\times 10^6$~events).
The central moments calculated for fixed $N_w$ generally decrease as we go towards higher $y$. In contrast to that, the curves
for $\mu_1$, $\mu_2$, and $\mu_4$ simulated with Glauber MC show peaks around $y = y_m$, i.e., where the antiprotons 
die out. Note that the antiproton rapidity distribution is cut more sharply than that of wounded protons. This results in a dip in 
both $S\sigma$ and $\kappa\sigma^2$as functions of $y$ around $y =1$.

Since $y_m$ increases with the collision energy, we would also expect that the position of the dip 
would move to higher  values of $y_m$. 
\begin{figure}[t]
\centerline{\includegraphics[width=0.33\textwidth]{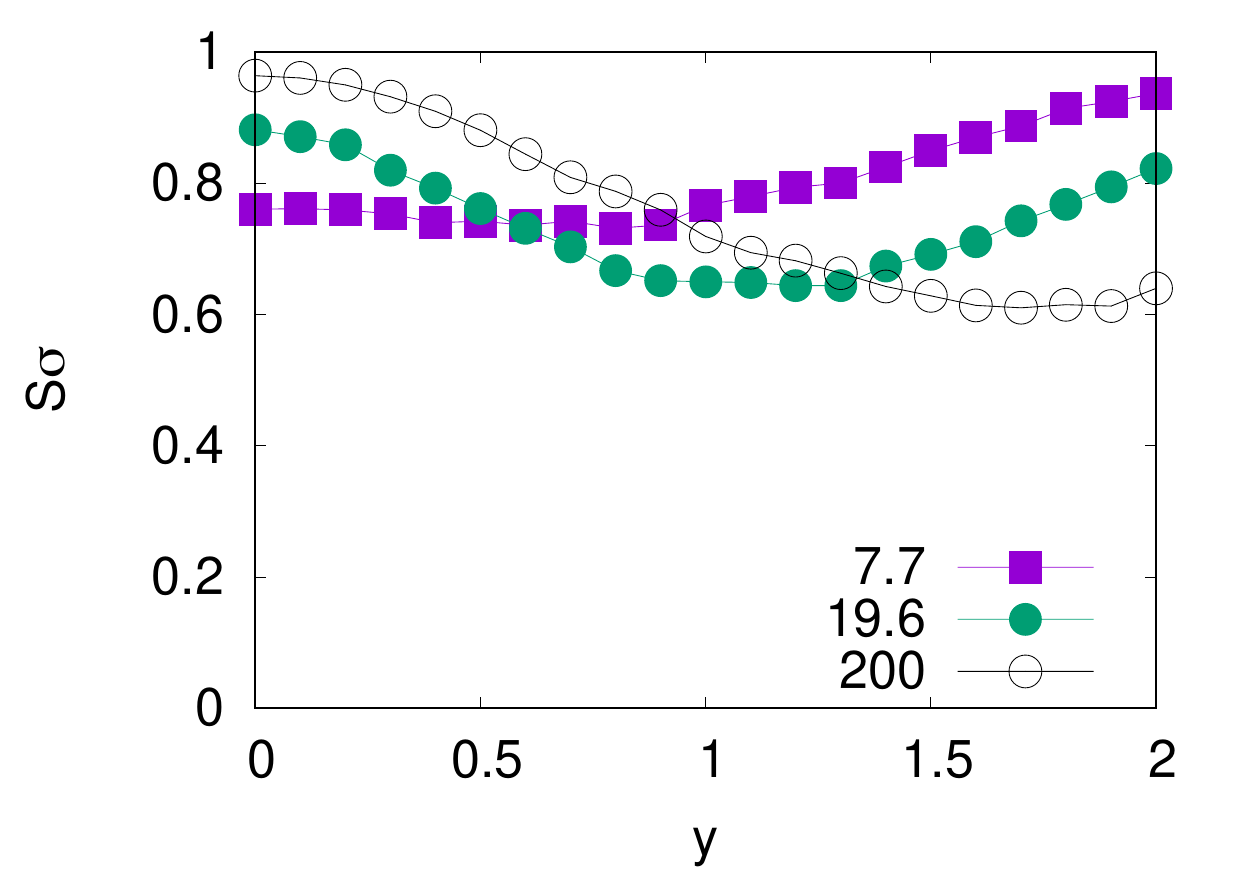}
\includegraphics[width=0.33\textwidth]{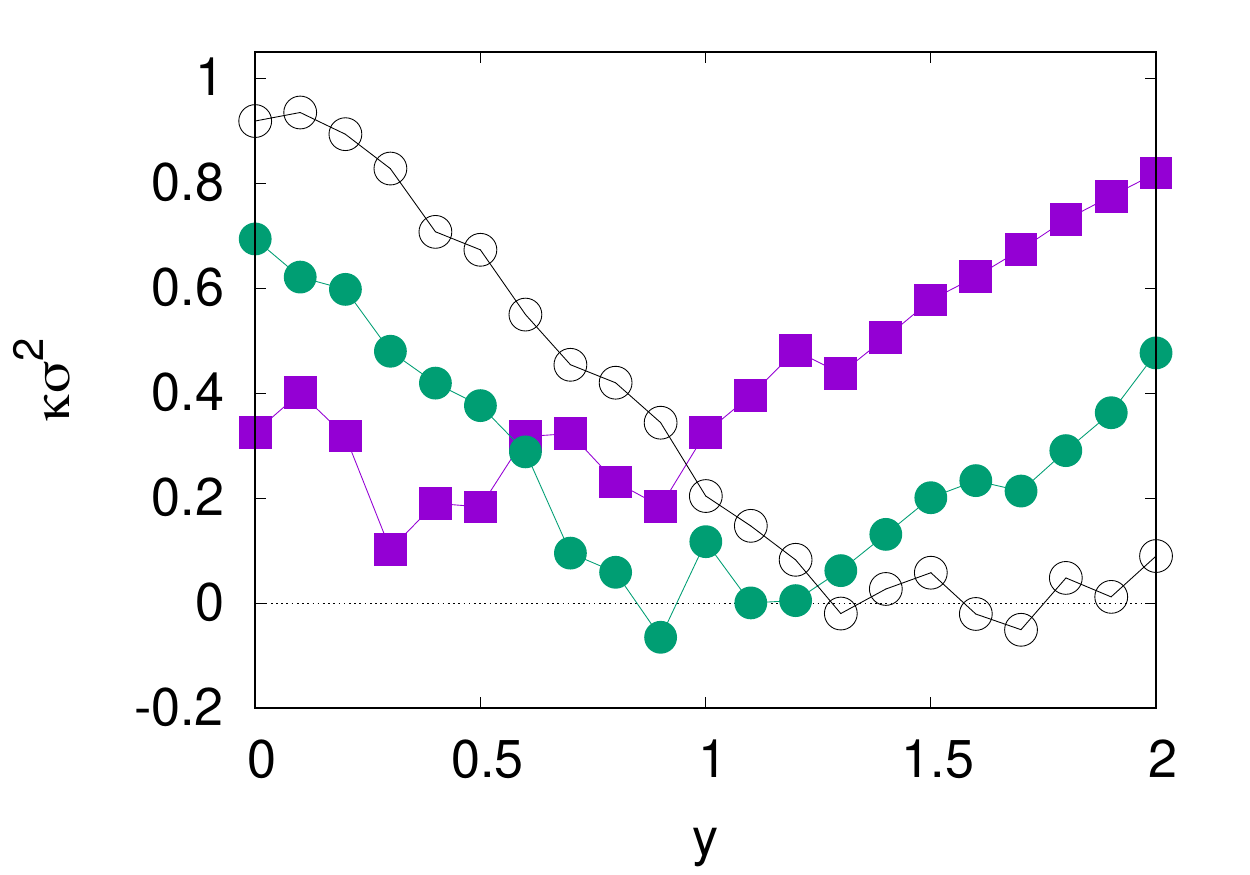}
\includegraphics[width=0.33\textwidth]{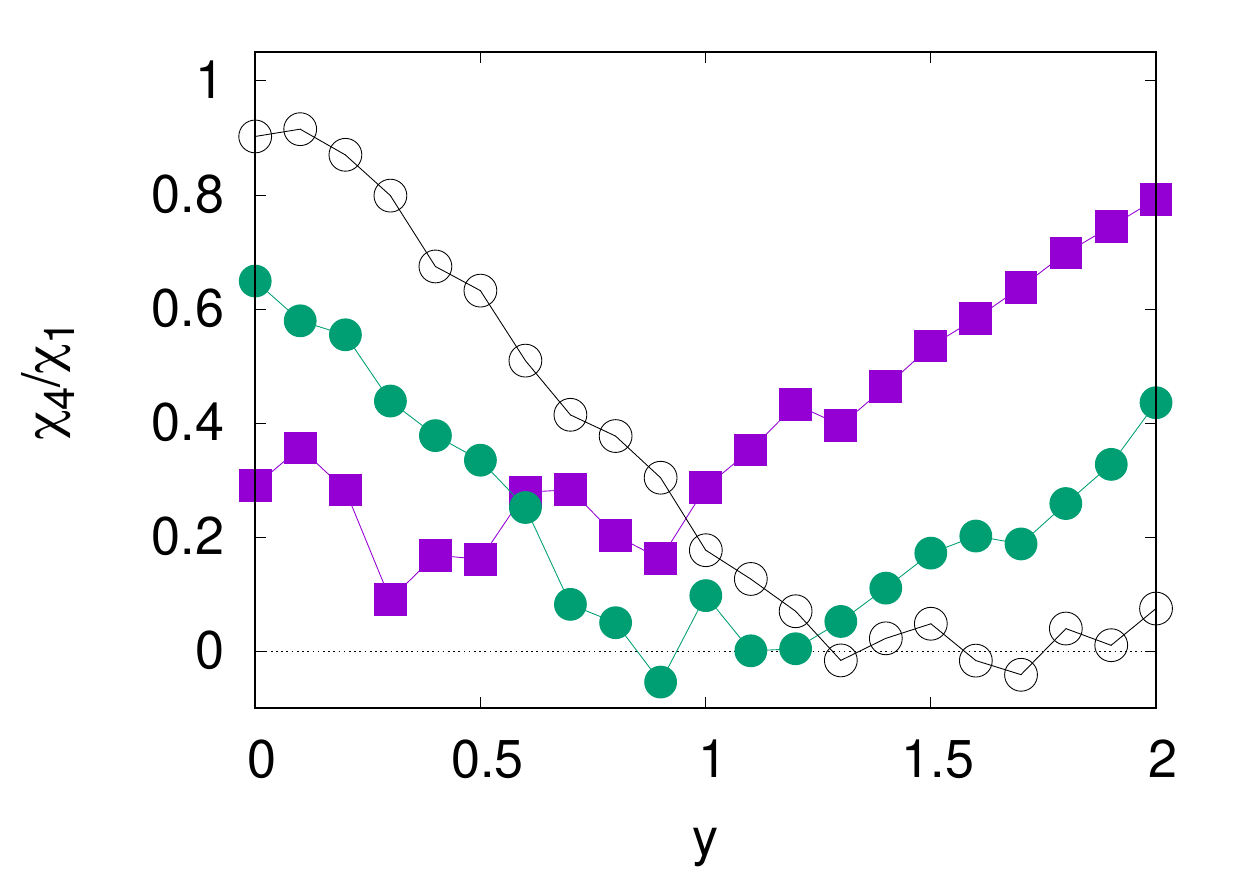}}
\caption{
Rapidity dependence of central moments, $S\sigma$, and $\kappa\sigma^2$ for three different collision energies, as indicated in the legend. 
For each energy $1.2\times 10^6$ events in 0--5\% centrality class generated by GLISSANDO have been analysed.
}
\label{f:YEdep}
\end{figure}	
This is confirmed in Figure~\ref{f:YEdep}, where we show $S\sigma$, $\kappa\sigma^2$, and $\chi_4/\chi_1$ as functions 
of $y$ for three collision energies. 
The parameters used in the simulations for all energies are summarised in Table~\ref{t:params}.
\begin{table}[b]
\caption{Parameters of the model for different collision energies.
\label{t:params}}
\begin{center}
\begin{tabular}{|c|c|c|}
\hline
$\sqrt{s_{NN}}$ [GeV]  & $y_m$ &  $N_{B\bar B}$  \\
\hline
7.7 & 0.519 & 0.8265 \\
11.5 & 0.770 & 4.4790  \\
19.6 & 1.019 & 16.946  \\
27 &  1.128 & 27.1070  \\
39 & 1.308 & 44.4262  \\
62.4 & 1.384 & 75.2842 \\
200 & 1.665 & 177.794 \\
\hline
\end{tabular} 
\end{center}
\end{table}
In the Figure we only show results from Glauber MC version of our model. Unfortunately, the 
statistics of $1.2\times 10^6$ events is not sufficient to avoid large statistical errors in the curves for the fourth order. Nevertheless,
one can see that this simple baseline model predicts rather strong dependence on $y$ of the fourth order cumulant ratios which 
for $\sqrt{s_{NN}} = 200$~GeV start
around 1 and fall down all the way to 0 at $y = 1.2$. The decrease is less severe when the collision 
energy decreases. 

In order to present the centrality dependence of the fluctuations we turn back to $\sqrt{s_{NN}} = 19.6$~GeV. 
\begin{figure}[t]
\centerline{\includegraphics[width=0.495\textwidth]{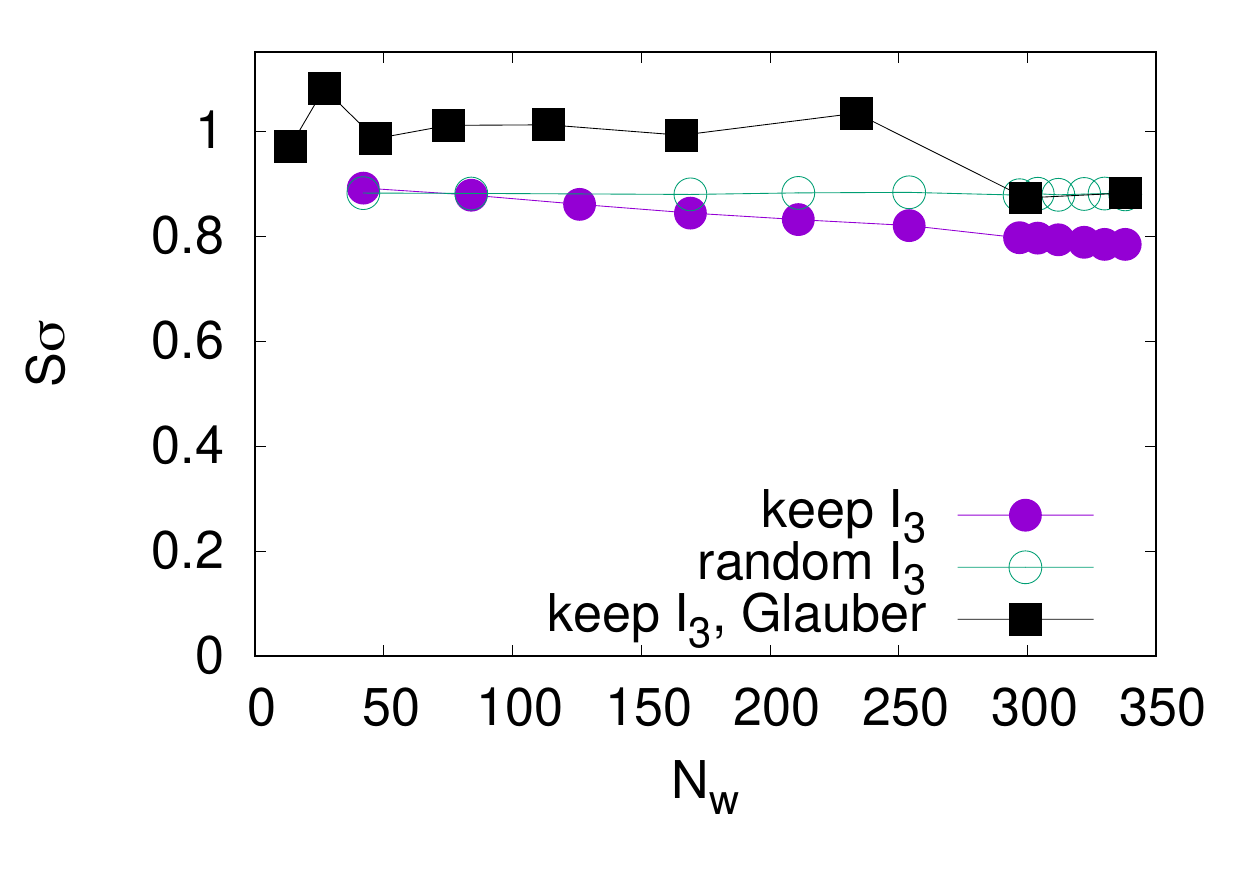}
\includegraphics[width=0.495\textwidth]{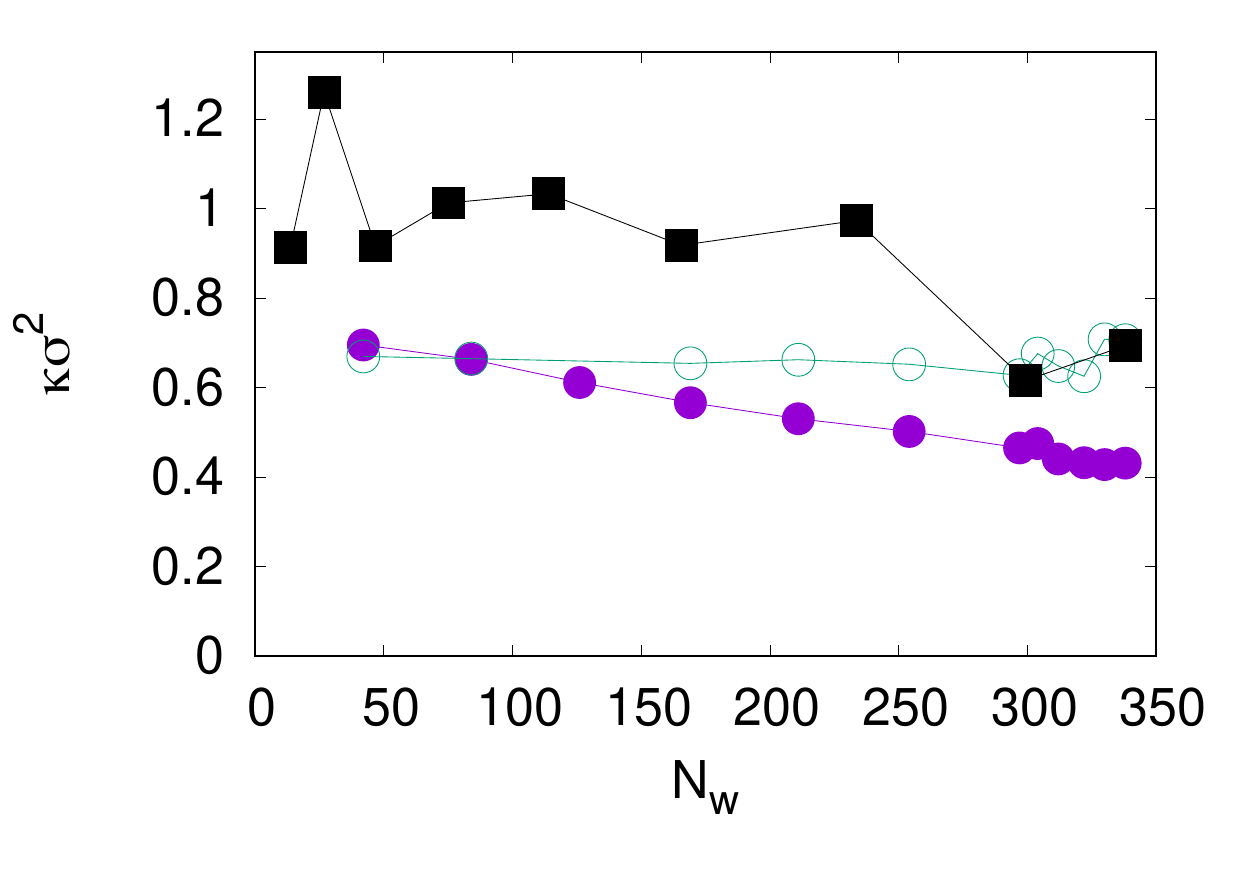}}
\caption{
Centrality dependence of $S\sigma$ and $\kappa\sigma^2$ for parameters corresponding to central Au+Au collisions at $\sqrt{s_{NN}} = 19.6$~GeV.
Full purple circles: fixed $N_w$ at values corresponding to mean at given centrality class  
and wounded nucleons remember their original isospin. Open green circles: fixed $N_w$ for given centrality class and 
wounded nucleons have their isospins randomised. Solid black squares: Glauber MC fluctuating $N_w$ corresponding to 0--5\% most central 
with wounded nucleons remembering their original isospin. 
}
\label{f:Cdep}
\end{figure}	
Selected results 
are shown in Fig.~\ref{f:Cdep}: $S\sigma$ and $\kappa\sigma^2$ as functions of the number of wounded nucleons $N_w$. 
The number $N_w$ is either fixed or we use for plotting the mean number from GLISSANDO for the selected centrality class. 
With these results we also show the influence of wounded nucleon isospin randomisation or isospin memory.  
We investigated the influence of this effect on the simulations with fixed $N_w$. The centrality dependence appears flat 
with randomised isospins. The effect of isospin memory is most pronounced in most central collisions since it is here that the 
limit on incoming proton number kicks in most strongly. As a result, the third and the fourth cumulant are slightly lowered in central 
collisions. The isospin memory has also been assumed in the simulations with the Glauber MC model. Although the statistics of 
$5\times 10^5$ events for each centrality still leads to large statistical uncertainties, we see that these fluctuations increase 
$S\sigma$ by about 0.1 and $\kappa\sigma^2$ by about 0.2.

\begin{figure}[t]
\centerline{\includegraphics[width=0.33\textwidth]{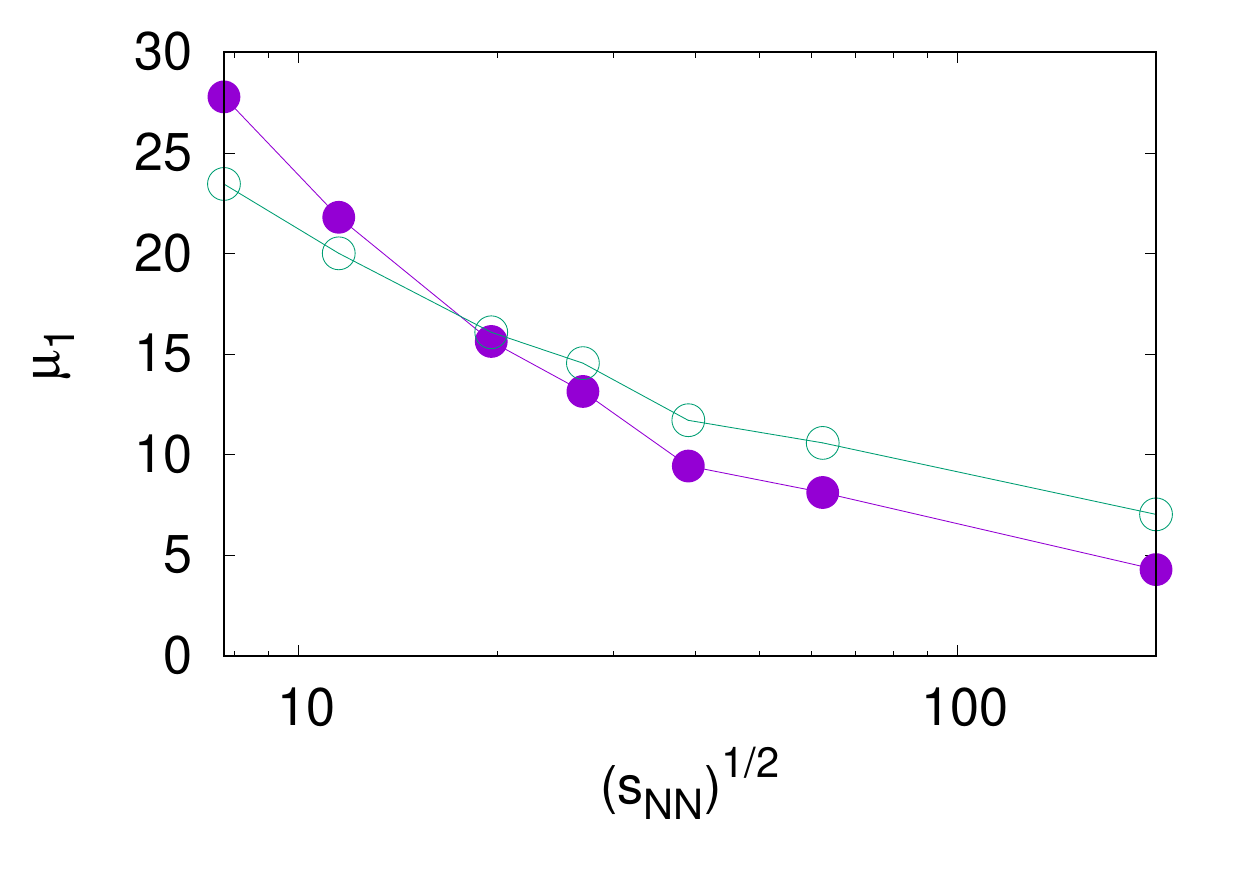}
\includegraphics[width=0.33\textwidth]{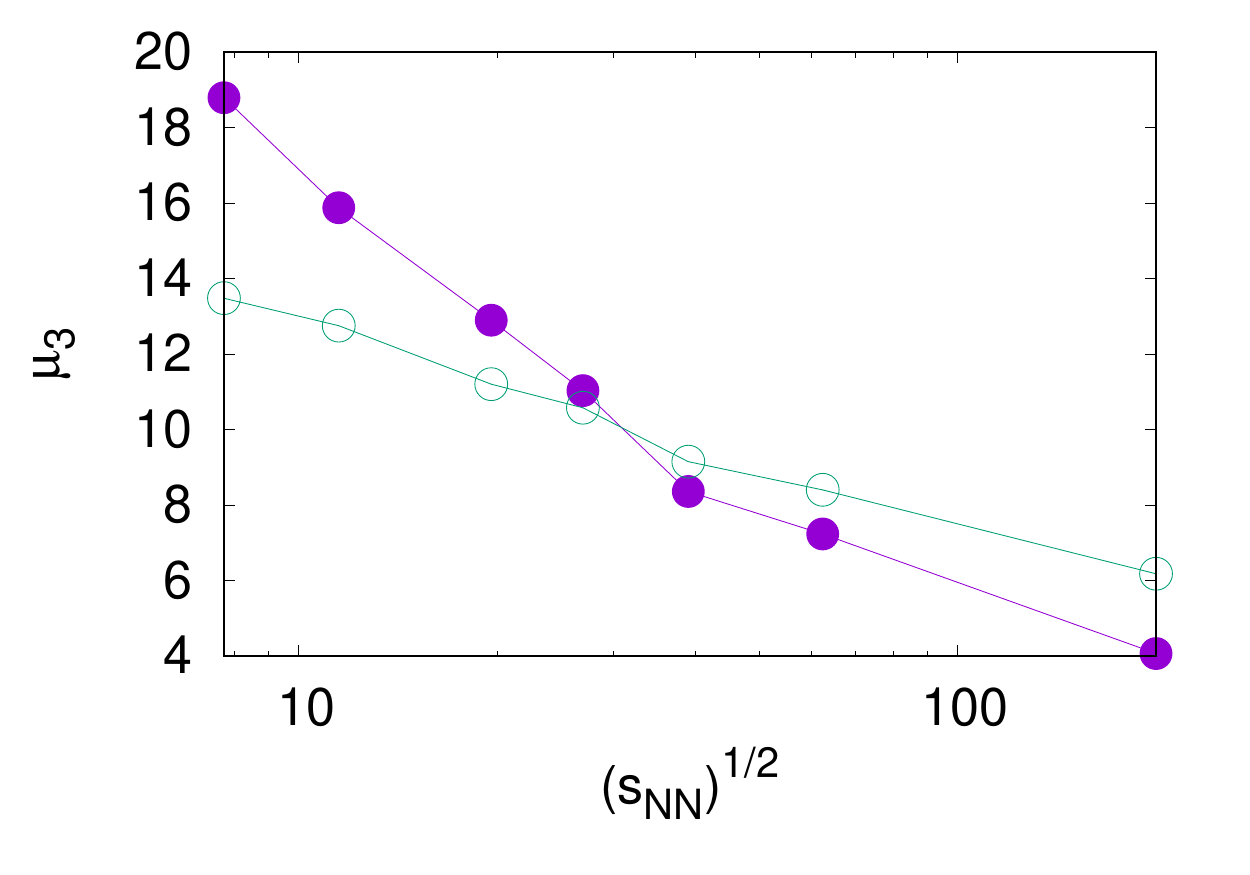}
\includegraphics[width=0.33\textwidth]{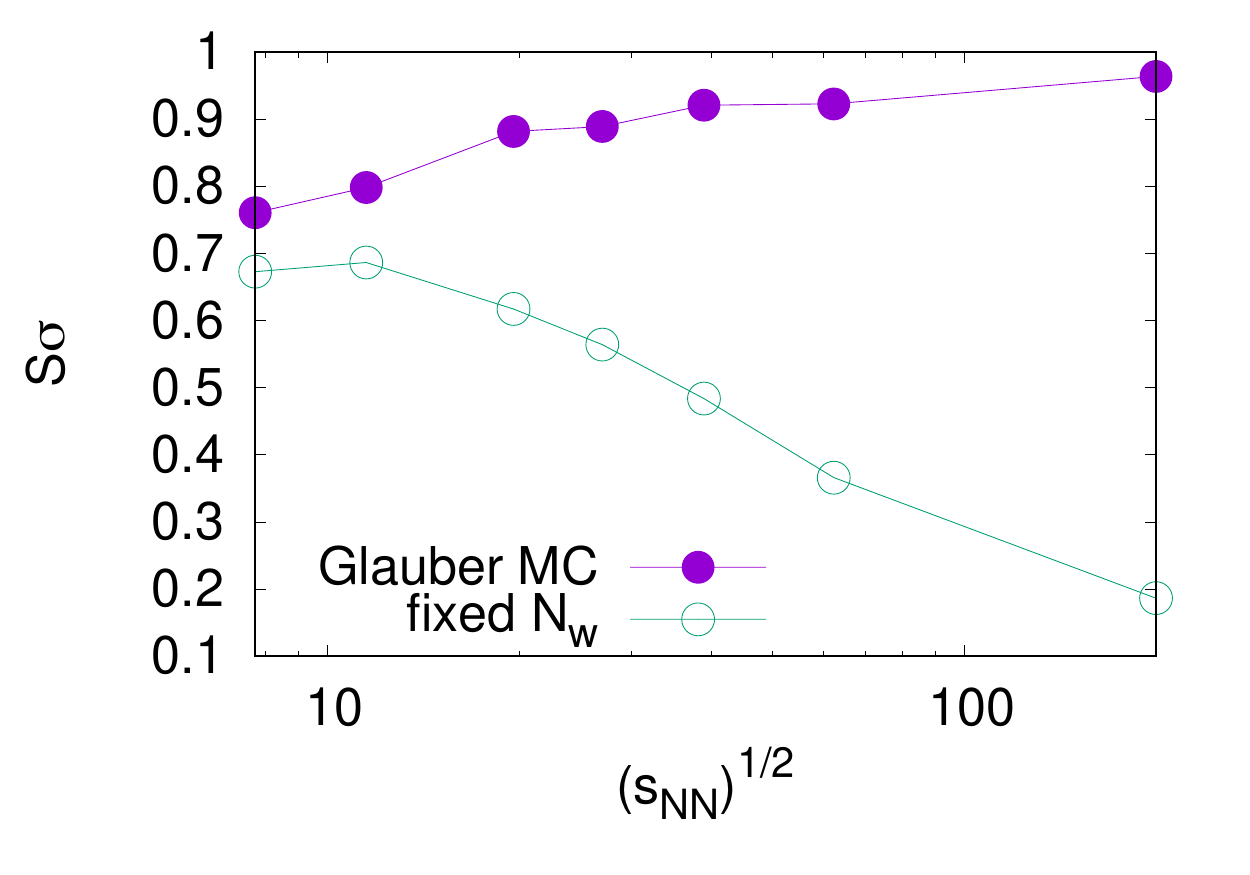}}
\centerline{\includegraphics[width=0.33\textwidth]{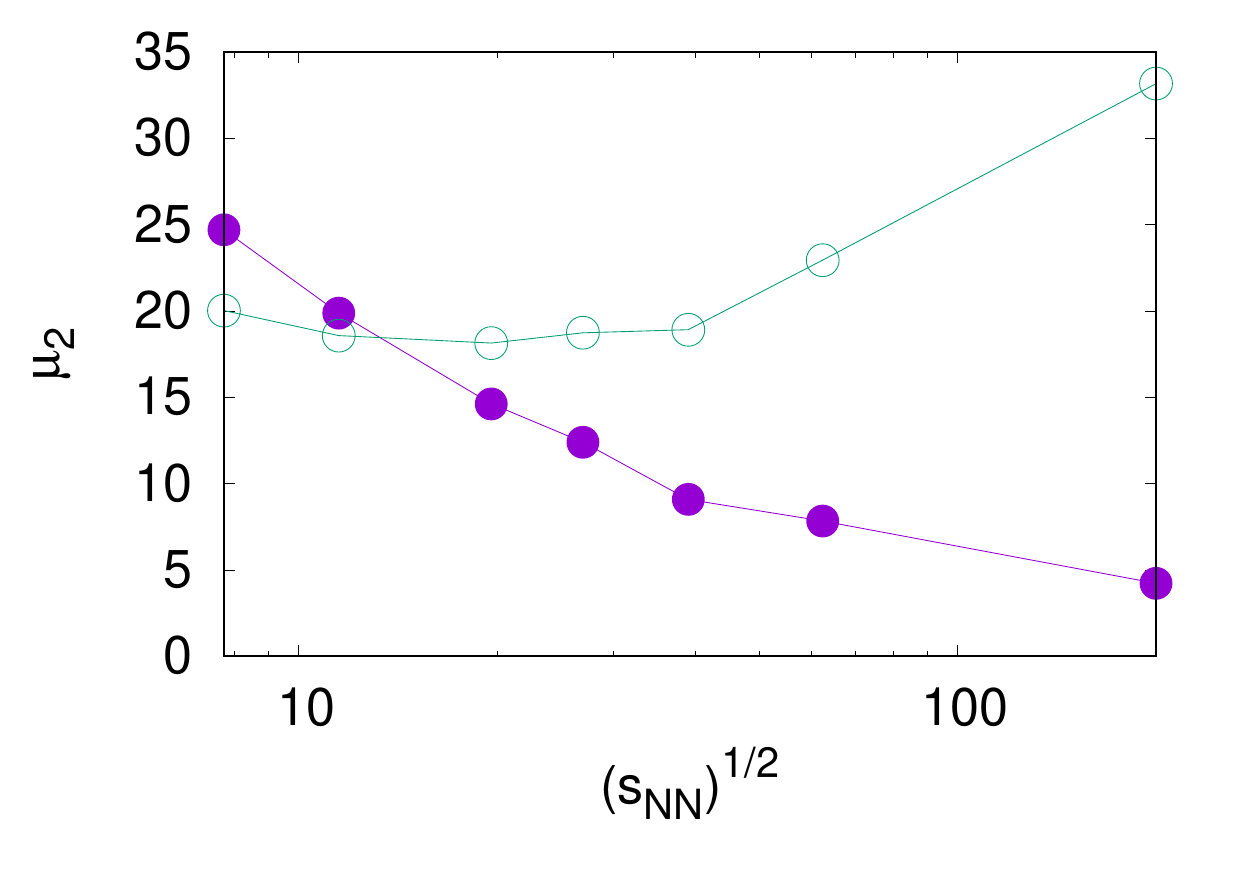}
\includegraphics[width=0.33\textwidth]{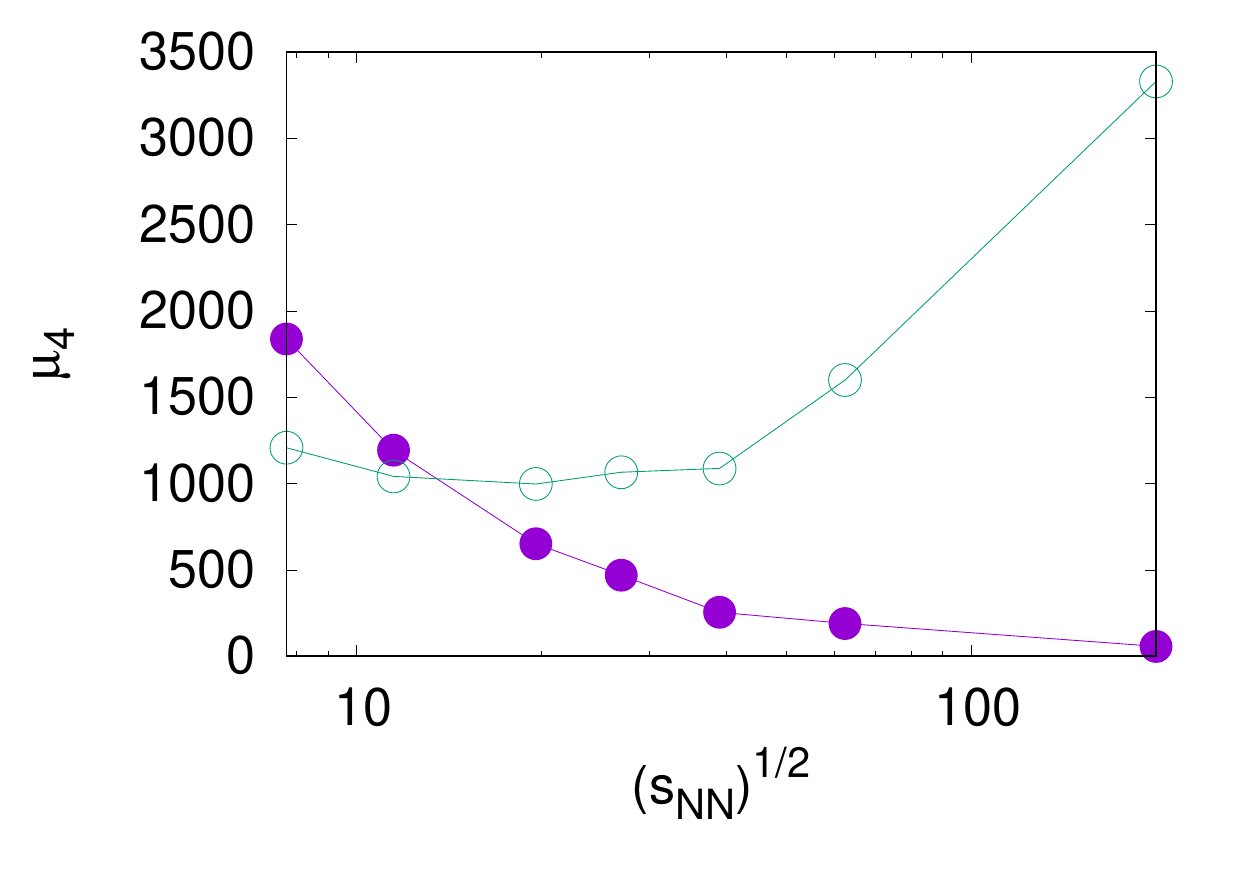}
\includegraphics[width=0.33\textwidth]{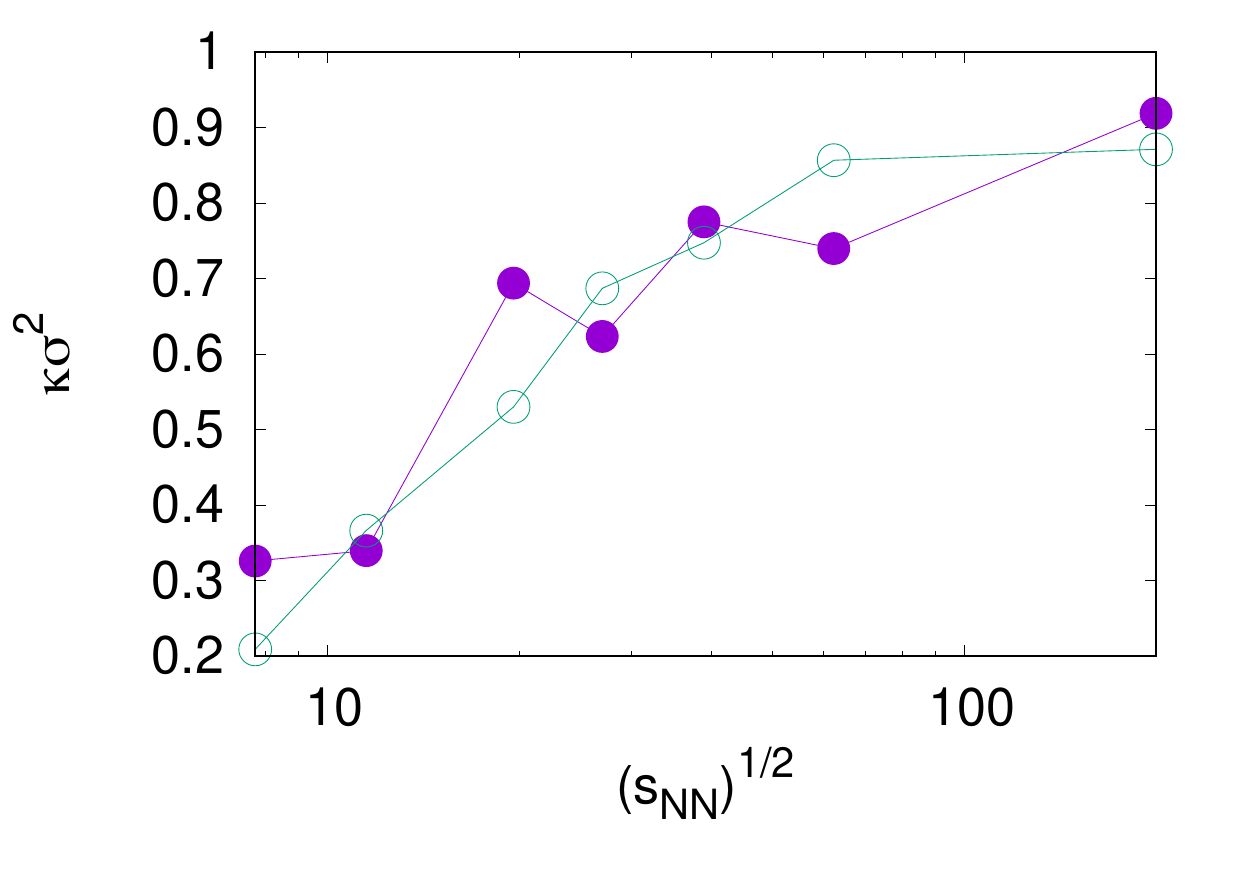}}
\caption{
Dependence of the central moments of the net=proton number distribution, $S\sigma$, $\kappa\sigma^2$ on $\sqrt{s_{NN}}$ in Au+Au collisions. 
Open green circles: simulations with fixed $N_w$. Full purple circles: simulations with 0--5\% most central collisions based 
on GLISSANDO Glauber MC. 
}
\label{f:Edep}
\end{figure}

Finally, we investigated the dependence of the moments on the collision energy in Figure~\ref{f:Edep}.
(Note that model parameters for all energies have been summarised in Table~\ref{t:params}.)
We show results for 0--5\% central  collisions. Since we are showing results for the central rapidity window
of $\Delta y = 0.5$, the baryon-antibaryon asymmetry decreases as the collision energy goes up. The relative 
contribution of produced $B\bar B$ pairs is increasing. This is demonstrated by $\sqrt{s_{NN}}$ dependence of 
$\mu_1$. In general, all higher central moments decrease, as well, if we look at the results obtained for fluctuating 
$N_w$. For $\mu_2$ and $\mu_4$ we see that the curves for fixed $N_w$ (i.e.~which do not include all baseline effects)
increase together with $\sqrt{s_{NN}}$. This behaviour is reflected in $S\sigma$ which starts  below 0.8 at low collision 
energies and approaches 1 as $\sqrt{s_{NN}}\to 200$~GeV for Glauber MC simulations, but is strongly decreased in 
fixed $N_w$ results.
The scaled kurtosis $\kappa\sigma^2$, on the other hand, does not change when $N_w$ fluctuations are turned on 
and off, since $\mu_2$ and $\mu_4$ behave similarly, here. The scaled kurtosis starts at very low values for 
$\sqrt{s_{NN}}=7.7$~GeV and grows together with $\sqrt{s_{NN}}$. Unfortunately, this is exactly opposite to the 
observed experimental result \cite{STARfluc} which shows a huge increase of $\kappa\sigma^2$ as 
$\sqrt{s_{NN}}\to 7.7$~GeV. Hence, our result just highlights the uncommonness of the measured value.


\section{Conclusions}

We have set up a benchmark model which includes many non-critical effects that influence the fluctuations of net-proton number. 
Generally, it is always necessary to look at such trivial predictions when interpreting the measured data. 

In particular, our model allowed to look at the dependence of the moments of the net-proton number distribution on the rapidity 
of the centre of the acceptance window. Our predictions, presented in Figures~\ref{f:Ydep} and \ref{f:YEdep}, are thus complementary 
to the proposal to use the rapidity dependence for looking for the critical point of the phase diagram \cite{Brewer}.

We have also looked at the effect of the wounded nucleons remembering their original isospin. Such a mechanism may be present
in nuclear collisions at $\sqrt{s_{NN}}$ of the order of a few GeV, because there the isospin randomisation may become ineffective.

Nevertheless, the presented baseline effects are not sufficient to explain all observed features of the data; in particular the huge 
enhancement of $\kappa\sigma^2$ towards $\sqrt{s_{NN}}=7.7$~GeV is not reproduced by this model. 
One has to, however, always keep them in mind when interpreting any measured result.


\subsubsection*{Acknowledgements}
This work was supported by the grant 17-04505S of the Czech Science Foundation (GA\v CR)
and the grant VEGA  1/0348/18 (Slovakia).
International collaboration was supported by collaboration grant in framework of the German-Slovak PPP programme and 
the COST Action CA15213 THOR.


\end{document}